%% file: main.tex
\documentclass{article}

\usepackage{PRIMEarxiv}

\usepackage[utf8]{inputenc} 
\usepackage[T1]{fontenc}    
\usepackage{hyperref}       
\usepackage{url}            
\usepackage{booktabs}       
\usepackage{amsfonts}       
\usepackage{nicefrac}       
\usepackage{microtype}      
\usepackage{lipsum}
\usepackage{fancyhdr}       
\usepackage{graphicx}       
\input{body/utils/used-packages.tex}
\input{body/utils/commands.tex}
\graphicspath{{images/}}     

\input{body/header/title.tex}

\input{body/header/author.tex}

\begin{document}
\maketitle

\begin{abstract}
\input{body/header/abstract.tex}

\end{abstract}


\input{body/sections/a-introduction.tex}

\input{body/sections/b-background.tex}

\input{body/sections/c-hk-poly.tex}
\input{body/sections/d-hkpoly-requirements-and-stakeholders.tex}

\input{body/sections/e-hkpoly-architecture.tex}
\input{body/sections/j-hkpoly-implementation-scenario.tex}

\input{body/sections/j.1-scenario.tex}

\input{body/sections/j.2-hkpoly-implementation-details.tex}

\input{body/sections/j.4-evaluation-considering-this-scenario.tex}

\input{body/sections/n-related-work.tex}
\input{body/sections/z-conclusion.tex}
\input{body/sections/zz-appendix}


\bibliographystyle{apalike-sol}
\bibliography{main}

\end{document}

%% file: body/utils/used-packages.tex
\usepackage[inline]{enumitem}
\usepackage{listings}
\usepackage{url}
\usepackage[normalem]{ulem}
\hypersetup{
    colorlinks=true,       
    linkcolor=red,          
    citecolor=blue,        
    filecolor=cyan,         
    urlcolor=blue        
}

%% file: body/utils/commands.tex
\newcommand{\eg}{\textit{e.g.,}}

\newcommand{\ie}{\textit{i.e.,}}
\newcommand{\etal}{\textit{et al.}}
\newcommand{\hk}{\texttt{HK}}
\newcommand{\hkplatform}{\texttt{HK\--Plat\-form}}
\newcommand{\hkpoly}{\texttt{HKPo\-ly}}
\newcommand{\hkpolyservice}{\hkpoly{} \texttt{ser\-vi\-ce}}
\newcommand{\hkbase}{\texttt{HKBa\-se}}
\newcommand{\kes}{\texttt{KES}}
\newcommand{\hyql}{\texttt{HyQL}}
\newcommand{\hkpolycomponent}[1]{\textit{#1}}
\newcommand{\hkpolycatalog}{\hkpolycomponent{HKPoly Knowledge Graph}}

\newcommand{\hkpolyclient}{\hkpolycomponent{Client Application}}

\newcommand{\hkpolyprovmanager}{\hkpolycomponent{HKProvManager}}

\newcommand{\dmelement}[1]{\texttt{#1}}
\newcommand{\dmelementAttribute}{\dmelement{At\-tri\-bu\-te}}

\newcommand{\dmelementComplexAttribute}{\dmelement{Com\-plex\-At\-tri\-bu\-te}}
\newcommand{\dmelementDatabase}{\dmelement{Da\-ta\-ba\-se}}

\newcommand{\dmelementDataStore}{\dmelement{Da\-ta\-Sto\-re}}

\newcommand{\dmelementDataSchema}{\dmelement{Da\-ta\-set\-Sche\-ma}}
\newcommand{\dmelementDatabaseSchema}{\dmelement{Da\-ta\-ba\-se\-Sche\-ma}}

\newcommand{\dmelementFileSystem}{\dmelement{Fi\-le Sys\-tem}}

\newcommand{\dmelementSchema}{\dmelement{Sche\-ma}}
\newcommand{\dmelementeg}[1]{\textit{#1}}

\newcommand{\ihkpoly}{\hkpolycomponent{IHKPoly}}
\newcommand{\hkdatasource}{\hkpolycomponent{HKDataSource}}
\newcommand{\postgresqlfdw}{\hkpolycomponent{PostgreSQL FDW}}

\newcommand{\buildingquery}{\texttt{BuildingQuery}}
\newcommand{\fdwqueryexecution}{\texttt{FDW Query Execution}}



%% file: body/header/title.tex
\title{A Polystore Architecture Using Knowledge Graphs to Support Queries on Heterogeneous Data Stores
\thanks{\textit{For authors marked with \textsuperscript{*}: \textbf{Work done while at IBM Research}}} 
}

%% file: body/header/author.tex
\author{
  Leonardo Guerreiro Azevedo \\
  IBM Research \\
  Rio de Janeiro, RJ, Brazil \\
  \texttt{lga@br.ibm.com}
\And
  Renan Francisco Santos Souza\textsuperscript{*} \\
  National Center for Computational Sciences \\
  Oak Ridge National Laboratory \\
  Oak Ridge, TN, USA \\
  \texttt{souzar@ornl.gov}
\And
  Elton F. de S. Soares, Raphael M. Thiago \\
  IBM Research \\
  Rio de Janeiro, RJ, Brazil \\
  \texttt{eltons@ibm.com, raphaelt@br.ibm.com}
\And
  Julio Cesar Cardoso Tesolin\textsuperscript{*} \\
  PGED/IME \\
  \texttt{jcctesolin@ime.eb.br} \\
\And
  Anna C. Oliveira\textsuperscript{*} \\
  PESC/COPPE \\
  \texttt{acoliveira@cos.ufrj.br} \\
\AND
  Marcio Ferreira Moreno\textsuperscript{*} \\
  MOBR Systems \\
  Sao Paulo, Brazil \\
  \texttt{marcio@mobr.ai}
}

%% file: body/header/abstract.tex
Modern applications commonly need to manage dataset types composed of heterogeneous data and schemas, making it difficult to access them in an integrated way.
A single data store to manage heterogeneous data using a common data model is not effective in such a scenario, which results in the domain data being fragmented in the data stores that best fit their storage and access requirements (\eg{} NoSQL, relational DBMS, or HDFS). 
Besides, organization workflows independently consume these fragments, and usually, there is no explicit link among the fragments that would be useful to support an integrated view.
The research challenge tackled by this work is to provide the means to query heterogeneous data residing on distinct data repositories that are not explicitly connected. 
We propose a federated database architecture by providing  a single abstract global conceptual schema to users, allowing them to write their queries, encapsulating data heterogeneity, location, and linkage by employing:
(i) meta-models to represent  the  global conceptual schema,  the remote data local conceptual schemas,  and  mappings  among  them;
(ii) provenance to create explicit links among the consumed and generated data residing in separate datasets.
We evaluated the architecture through its implementation as a polystore service, following a microservice architecture approach, in a scenario that simulates a real case in Oil \& Gas industry.
Also, we compared the proposed architecture to a relational multidatabase system based on foreign data wrappers, measuring the user's cognitive load to write a query (or query complexity) and the query processing time. 
The results demonstrated that the proposed architecture allows query writing two times less complex than the one written for the relational multidatabase system, adding an excess of no more than 30\% in query processing time.

%% file: body/sections/a-introduction.tex
\section{Introduction}
\label{sec:introduction}

Several modern applications manipulate diverse datasets with different models and usages, employing specific tools and techniques, \eg{} applications in medical informatics, oceanography, metagenomics, and exploration and production phases in Oil \& Gas~\cite{souza2019escience}. 

\begin{figure}[tb]
    \includegraphics[width=0.9\textwidth]{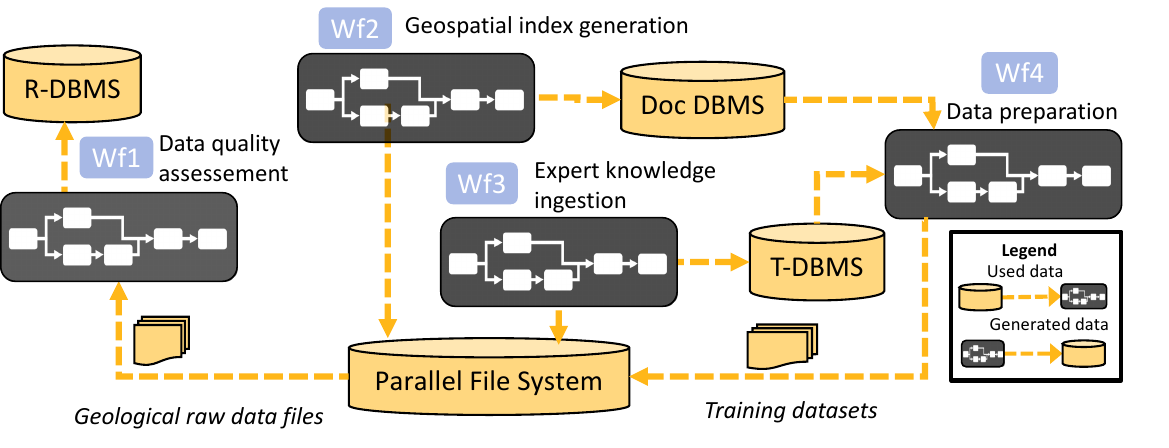}
    \centering
    \caption{Workflows and  data stores of the oil reserves  discovery  scenario (adapted from~\cite{souza2019escience}).}
    \label{fig:oil-and-gas-wokflow-scenario}
\end{figure}

For example, in an oil reserves discovery scenario, independent workflows process sizeable raw data files to generate training and validation datasets used by Deep Learning (DL) models, as illustrated in Figure~\ref{fig:oil-and-gas-wokflow-scenario}. Although the workflows do not have direct relationships, they comprise several activities that consume and generate data from/to heterogeneous data stores in which data are not directly linked.

The first workflow processes geological raw data files to extract necessary metadata.
It also assesses missing and displaced information as a measurement of metadata quality.
While data files reside on a Parallel File System, the data quality score and extracted metadata are stored in a relational DBMS (R-DBMS). 
The second workflow considers the high-quality data files and generates geospatial indexes.
This generated data is used to accelerate geospatial queries over the geological data and is stored in a document-oriented DBMS (Doc DBMS). 
The third workflow also considers the high-quality data files and augments the raw geodata files with extra knowledge informed by geoscience experts. This meta-information is stored in a Triplestore System (T-DBMS). Finally, the last workflow prepares the learning datasets used by DL algorithms. 
It queries the data from Doc DBMS and the related knowledge from the T-DBMS to generate datasets, which are stored as HDF5 files in the Parallel File System, ready to be used by DL algorithms.

In this scenario, the user is a Machine Learning (ML) expert with deep knowledge in the domain. 
When reporting the ML model, the user must query the data residing in the heterogeneous data stores.
For instance, he/she should run a query to discover \textit{what are the geographic coordinates extracted from the SEG-Y file that is being used to produce training and validation files and the spatial resolution between slices in the seismic data}.
To answer this query, he/she has to inspect data residing on the file and the document database.
Besides using specific tools to access both systems, the user should also know how the data is inter-related,  \ie{} which lines of the file relate to objects of the document database.
These issues illustrate the challenges of handling such scenarios:
(i) the access to heterogeneous data stores;
(ii) the linkage of heterogeneous data residing in remote distinct data stores.

The research question addressed in our work can be shortly enunciated as \textit{How to query data that are not explicitly connected residing on heterogeneous data stores?}

Several data management solutions have emerged to handle heterogeneous data access, such as 
    distributed file systems (\eg{} GFS~\cite{ghemawat:2003:gfs} and HDFS\footnote{\url{https://hadoop.apache.org/docs/r1.2.1/hdfs_design.html}}), 
    NoSQL databases (\eg{} MongoDB\footnote{\url{https://www.mongodb.com/}}, AllegroGraph\footnote{\url{https://allegrograph.com/}}),
    new data processing frameworks (\eg{} Spark\footnote{\url{https://spark.apache.org/}}), and
    hybrid multimodal (\eg{} OrientDB\footnote{\url{https://orientdb.org/}}) or hybrid NewSQL (\eg{} LeanXcale\footnote{\url{https://www.leanxcale.com/}}).
In general, each solution handles a few kinds of data models or formats and does not addresses the diverse datasets required by modern applications~\cite{azevedo-et-al:2020:modern-federated-ds-overview}.
On the other hand, migrating the heterogeneous data to a single database and schema does not work. 
Data conversion and loading is a very costly and time-consuming task, which is difficult to justify, and it may involve the integration of disparate data (\eg{} integration of structured data with text, web pages, semi-structured data, time series). Besides, no DBMS offers high performance in all kinds of data, \ie{} one size does not fit all~\cite{stonebraker:2015:polystore}.

Therefore, a middleware that provides a seamless interface with an independent data model and (perhaps) data schemes is required to access heterogeneous data stores~\cite{stonebraker:2015:polystore}.
A Multidatabase System (MDBS) (or federated data system) provides a software layer that runs on top of individual autonomous data systems (or data stores) that share portions of their data. 
The MDBS allows users to access the various databases in an integrated way.
A polystore system, a kind of MDBS, provides integrated access to several data stores, such as NoSQL, RDBMS,  or  HDFS, sometimes through a  data processing framework~\cite{ozsu-valduriez:2020:principles-of-data-systems}.
Nevertheless, there is still a lack of abstraction and query language to access heterogeneous data stores in an integrated way~\cite{stonebraker:2015:polystore}.

To answer our research question, we propose a federated data system architecture that not only encapsulates heterogeneous data sources but also uses provenance for data linkage, \ie{} we create the explicit links by collecting the provenance of the workflow activities execution. 
Provenance contains information about the process and data used to derive the data product. We use causality as the kind of provenance information, which is the dependency relationships among data products and the processes that generate them, such as data-process dependencies and data dependencies~\cite{davidson-and-freire2008provenance}.
So, the architecture provides an abstract layer to users to formulate queries based on a global conceptual schema, and, to be able to process the user query, it supports the creation of data mappings and record linkage transparent to service consumers.
Its features are for:
    (i) Creation of global conceptual schema (the domain schema);
    (ii) Creation of local conceptual schemas for each external heterogeneous data store;
    (iii) Creation of mappings between global and local conceptual schemas;
    (iv) Creation of workflows' provenance schema;
    (v) Instrumentation of the applications that run the workflows;
    (vi) Capture of provenance during workflows execution;
    (vii) Processing of user queries.

We demonstrate the proposed architecture through its implementation as a RESTful web service in a microservice architecture~\cite{richardson-and-ruby:2008:restful-web-services}.
We devise models to represent the global conceptual schema, local conceptual schemas, and mappings among them besides provenance schema and execution.
This information is stored in a Knowledge Graph created using Hyperknowledge (\hk{}), a hybrid conceptual model able to handle the multitude of media content and the meaning behind this data~\cite{moreno-et-al:2017:extending-hypermedia-to-hyperknowledge}.
In particular, we are using the \textit{context} element of \hk{} to encapsulate and modularize the models and ingested data.

We evaluate the architecture implementation in a scenario that simulates a real case in Oil \& Gas industry depicted previously in this section.
The evaluation considered query complexity from the user and system perspectives. 
For the former, we analyzed the query components (like keywords, involved entities), and for the latter, we analyzed query execution time.
We analyzed the architecture through comparing it against PostgreSQL FDW (Foreign Data Wrapper)\footnote{\url{https://www.postgresql.org/docs/current/postgres-fdw.html}}. 
PostgreSQL is a widely used commercial open-source database management system and 
FDW is a module that allows access to external data stores using SQL language and wrappers.
Our evaluation results demonstrated that while the query is more than two times less complex than the same written in SQL, the overhead created in query processing time is not greater than 30\%.
It indicates the applicability and utility of the architecture; although further investigations and improvements are required in query processing time.

The remainder of this work is divided as follows.
Section~\ref{sec:background} presents the background and justify our choices.
Section~\ref{sec:hk-poly} presents the proposed architecture.
Section~\ref{sec:implementation} presents the proposal implementation and its evaluation.
Section~\ref{sec:related-work} presents the related work.
Section~\ref{sec:conclusion} presents the conclusion and proposals of future work.

%% file: body/sections/b-background.tex
\section{Background}
\label{sec:background}

This section presents the main concepts related to this work and the reasoning for their use in our proposal.

\subsection{Multidatabase and Polystore system}
\label{sec:multidatabase}

A Multidatabase System (MDBS) provides a layer of software that runs on top of individual Database Management Systems (DBMSs) and facilitates users'  access to various databases.
The MDBS presents a Global Conceptual Schema (GCS) to the user, representing an integrated view of a database in which parts are allocated to different sites.
In the MDBS environment, each site runs an individual DBMS, sharing some of its parts with the MDBS. 
Each local database is represented by a Local Conceptual Schema (LCS)~\cite{ozsu-valduriez:2020:principles-of-data-systems}.
The database consumer formulates queries based on the GCS, and 
the MDBS translates them into a group of local queries and sends them to be executed by the individual DBMSs.
The MDBS receives the responses, consolidates the queries' results, and returns an integrated result to the user~\cite{ozsu-valduriez:2020:principles-of-data-systems}.

\begin{figure}[tb]
    \centering
    \includegraphics[width=0.6\textwidth]{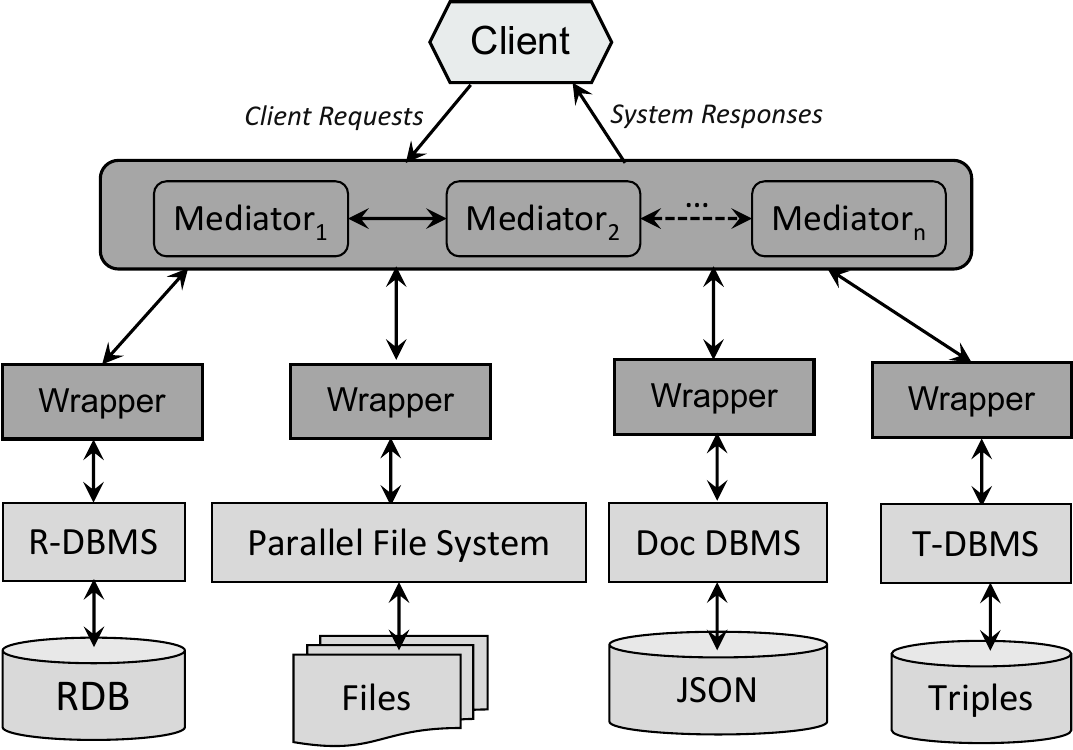}
    \caption{Layered Mediator/Wrapper Multidatabase System architecture (adapted from ~\cite{ozsu-valduriez:2020:principles-of-data-systems}).}
    \label{fig:mediator-wrapper-mdbs}
\end{figure}

The Mediator/Wrapper\footnote{
A mediator ``is a software module that exploits encoded knowledge about certain sets or subsets of data to create information for a higher layer of applications'' (\cite{wiederhold:1992:mediators} \textit{apud} \cite{ozsu-valduriez:2020:principles-of-data-systems}).}
is a kind of MDBS implementation.
It uses a common GCS and interface language(s).
The wrappers handle the heterogeneity performing mappings between LCS and GCS.
Figure~\ref{fig:mediator-wrapper-mdbs} illustrates this architecture with a mediator layered implementation.
The wrappers encapsulate parts of the global database provisioned by heterogeneous data sources (\eg{} Relational-DBMS, Parallel File System, Document DBMS, and Triplestore).

The MDBS definition can be specialized when the encapsulated systems include other data store systems besides DBMS.
Ozsu and Valduriez define this system as a polystore system, and it provides integrated access to several data stores, such as NoSQL, RDBMS, or  HDFS,  sometimes through a  data processing framework (Figure~\ref{fig:mediator-wrapper-mdbs}). 
Typically, they support only read-only queries,  as processing distributed transactions across data stores is a hard problem~\cite{ozsu-valduriez:2020:principles-of-data-systems}. 
We are proposing a polystore architecture in this work.

\subsection{Postgres FDW and SQL/MED}
\label{sec:postgres-fdw-and-slq-med}
The International Organization for Standardization (ISO) developed and published the SQL/MED (Management of External Data)\footnote{\url{https://www.iso.org/standard/63476.html}}, an extension of SQL that allows applications to use standard SQL to access SQL and non-SQL data concurrently. 
SQL/MED defines an API that standardizes the communication between a SQL-based Server and wrappers (aka, Foreign-Data Wrappers or FDW) to access external servers (aka Foreign Servers) data.
In other words, SQL/MED defines structures and routines the SQL-based Server and Foreign-Data Wrappers must provide \cite{melton:2001:sql-med}\cite{melton:2002:sql-med-status-report}.
Therefore, SQL/MED allows Relational-DBMS to work as an MDBS in a standard way, providing a homogeneous query language (SQL) and heterogeneous data structured (relational tables) to users.
PostgreSQL supports SQL/MED since version 9.1\footnote{\url{https://wiki.postgresql.org/wiki/SQL/MED}}. 
MySQL and MariaDB also implement the standard.
Other DBMSs do not implement the standard, although they provide similar implementations to access heterogeneous data.
Db2 provides distributed database feature, 
Microsoft SQL-Server offers Linked Server and 
Oracle provides database link~\cite{otuonye2021cloud-erp-smes}.

There are several foreign data wrappers already implemented for PostgreSQL\footnote{\url{https://wiki.postgresql.org/wiki/Foreign_data_wrappers
}}, encapsulating systems like SQL DBMS, NoSQL Databases, File Systems, Scientific tools, Operating Systems, etc.
Besides, many works published in the literature use PostgreSQL FDW in their proposals. 
For example, more than 25 papers are returned using the search string \texttt{("foreign data wrapper" OR fdw) AND (Postgres or PostgreSQL)} in IEEE and ACM digital libraries.
The robustness of PostgreSQL DBMS, its open-source license, the wrappers already implemented, and its wide use in literature motivated us to employ PostgreSQL FDW in validating our proposal.

\subsection{Provenance}
\label{sec:provenance-provlake}

Provenance is also known as the audit trail, lineage, and pedigree of a data product. It contains information about the process and data used to derive the data product~\cite{davidson-and-freire2008provenance}.

In this work, we are employing ProvLake\footnote{\url{http://ibm.biz/provlake}}, a lineage data management system capable of capturing, integrating, and querying data across multiple workflows by leveraging provenance data. 
ProvLake's data model is a provenance data representation for workflows on data lakes \cite{souza2019escience} which is built on W3C PROV~\cite{W3CPROV} and PROV-Wf~\cite{costa-et.al.:2013:capturing}. 
ProvLake tracks data in distributed and heterogeneous environments. 

ProvLake provides a lightweight \textit{data tracking API} to be added to workflow codes (ProvLakeLib\footnote{\url{https://github.com/IBM/multi-data-lineage-capture-py}}), such as scripts. 
Also, it provides a \textit{query API} for runtime analytical queries that integrate multistore data. 
When combined with a polystore, it can query data directly in multiple stores jointly with their provenance data.
Conversely, our solution enables access to remote data not ingested during the capture process using remote data links loaded during provenance capture.

\subsection{Knowledge Graph and Hyperknowledge (\hk{})}
\label{sec:hyperknowledge}

Knowledge Graphs have gained a broad use in research and business~\cite{ji-etal:2021:survey-on-knowledge-graphs} since the term was coined by a Google blog post~\cite{singhal:2012:introducing-the-knowledge-graph} in 2016~\cite{ehrlinger-and-woss:2016:towards-a-definition-of-knowledge-graphs}. 
There are several definitions of the term in the literature, and 
after a terminological analysis and based on the typical architecture of a knowledge-based system - which has information sources and is composed of a knowledge base (\eg{} ontology\footnote{Gruber defined ontology as ``a set of representational primitives with which to model a domain of knowledge or discourse''~\cite{gruber:2008:ontology}.}) and a reasoning engine components, Ehrlinger and W{\"o}{\ss}'s proposed the following definition~\cite{ehrlinger-and-woss:2016:towards-a-definition-of-knowledge-graphs}:
\begin{quote}
    A knowledge graph acquires and integrates information into an ontology and applies a reasoner to derive new knowledge.
\end{quote}
We follow this definition in our work since we create a knowledge base acquiring information of schemas and instances corresponding to the GCS, LCS, mappings, and provenance, and we use inference mechanisms for, \eg{} to navigate in the KG and compute queries to be executed on the remote data stores. 
To accomplish it, we use the IBM Hyperlinked Knowledge Graph\footnote{\url{https://github.com/ibm-hyperknowledge}} (or Hyperknowledge (\hk{}) for short).

\hk{} is a hybrid conceptual model able to handle the multitude of media content and the meaning behind this data~\cite{moreno-et-al:2017:extending-hypermedia-to-hyperknowledge}. 
More specifically, it supports the description of associations among symbolic semantics and non-symbolic data fragments within the same knowledge base. 
This representational approach takes advantage of combining in a single rationale user interaction, data segments (\eg{} sentences of a text document, fragments of images, segments of seismic data, frames of a video file, fragments of executable code), and semantic representations (\eg{} knowledge entities in an ontology that can be reasoned upon)~\cite{moreno-et-al:2017:extending-hypermedia-to-hyperknowledge}. 

The foundation of \hk{} is a pair of typical hypermedia concepts:
    (i) Nodes: represent information units;
    (ii) Links: define relationships among fragments of information (anchors) and properties of nodes.

Nodes can be of two classes:
   (i) Terminal nodes are composed of a collection of information units\footnote{The exact notion of what constitutes an information unit is part of the node definition and depends on its specialization. For instance, a terminal node may represent a class (concept node) in an ontology or a media content (content node), such as a video~\cite{moreno-et-al:2017:extending-hypermedia-to-hyperknowledge}.};
   (ii) Composite nodes, whose content is a set of nodes of the two classes. 
Composite nodes may also be specialized to better define the semantics of node collections. 
For example, a \dmelement{Context} node is a composite node that contains a set of links and other attributes. 
They are handy for grouping knowledge and content specifications in logical containers~\cite{moreno-et-al:2017:extending-hypermedia-to-hyperknowledge}.

\subsection{Hyperknowledge Platform --- \hkplatform{}}
\label{sec:hkplatform}

\hkplatform{} is a set of tools developed for handling \hk{}. From this set, we use two of them: \hkbase{} and \kes{}.

\hkbase{} provides a RESTful API to manipulate structured data represented in \hk{} and unstructured data (\eg{} images, videos). 
Structured data may be stored and retrieved from single data stores of different types (\eg{} Triplestores - Jena or AllegroGraph, Graph Databases - JanusGraph, Document Databases - MongoDB) while unstructured data is stored in Object Stores or File Stores (\eg{} MinIO, FileSystem).

In \hkbase{}, data may be retrieved by using either \hyql{} (the Hyperknowledge Query Language) or SPARQL. 
\hyql{} is the language created to query data represented using \hk{} constructs~\cite{soares-et-al:2021:hk-dataset-engineering}.
On the other hand, SPARQL is the query language for RDF (Resource Description Framework) defined by W3C~\cite{sparql:2008:sparql-query-language-for-rdf}.
RDF is a directed, labeled graph data format where datasets are represented as triples (subject, predicate, object)~\cite{zulkefli-et-al:2013:evaluation-of-triple-indices}, like the triple \texttt{(LeonardoDaVinci, hasCreated, TheMonalisa}). 
\textit{Triplestores}, or RDF stores, are the matter of choice for storing and querying RDF data, like Jena and AllegroGraph. 
So, when \hkbase{} is storing \hk{} in a Triplestore database, \hkbase{} provides a solution to query the underlying \hk{} data using SPARQL, \ie{} one can query the database as if only RDF data was stored, filtering specific \hk{} constructs.

\kes{} (Knowledge Explorer System)~\cite{moreno:2018:kes} is a web application that provides a user interface (UI) for collaborative management of \hk{} bases (\ie{} knowledge bases with \hk{} specifications). 
\kes{} enables creating, validating, and curating \hk{} descriptions in an interactive visual approach. 

Our polystore solution is provisioned as an \hkbase{} service and
\kes{} is used to inspect the ingested data and as an alternative to manually manage models and mappings.

\subsection{Query complexity}
\label{sec:query-complexity}

The query complexity can be measured using the database and the user perspective.
While the former considers the required time and the number of resources to run a query, the latter examines the user's cognitive load to read and write a query. 
Although query complexity measurement methods are well established for the database perspective, a few works address the user perspective~\cite{vashistha-and-jain:2016:measuring-query-complexity}.

Ozsu and Valduriez~\cite{ozsu-valduriez:2020:principles-of-data-systems} state that the number of relations and operators characterizes complex SQL queries. 
The complexity increases with the increasing number of equivalent operators.

Vashistha and Jain~\cite{vashistha-and-jain:2016:measuring-query-complexity} measure query complexity by:     
    number of tables and columns in a query; 
    query length (\ie{} number of characters); 
    number of operators (\eg{} join, filter); 
    query expressions (\eg{} like, greater than, or, and);
    and query runtime.

Yu \etal{}~\cite{yu:2019:spider} evaluate SQL query complexity considering the number of query components, like keywords (\eg{} group by, order by, intersect), nested subqueries, column selections, and aggregators. 
They define weights for each component, considering that the use of some of them makes the query harder to understand than others. 
In the same direction, Subali \etal{}~\cite{subali:2018:measuring-sql-complexity} propose a method to evaluate SQL query complexity based on the number of occurrences of SQL components classified in Variable Output, Variable Input, Nested Query, Join Table, and Number of Tables. 
The query complexity score is calculated using the number of occurrences of each SQL component multiplied by the assigned weight of its category.

For the SPARQL query language, Yuanbo  \etal{}~\cite{guo:2005:lubm} measure query complexity by the number of classes and properties. 
They used this approach to evaluate the LUBM benchmark queries. 
Souza  \etal{}~\cite{Souza:2021:provml-extended} use a similar approach while evaluating workflow provenance techniques to build a holistic view to support the lifecycle of scientific ML. 
They measure query complexity based on the number of filter clauses, the patterns to match in the graph traversal, aggregations, and sorting, and the number of triples satisfying the patterns to match. 

In this work, both perspectives of query complexity concern us.
Our goal is to provide a model to the user that encapsulates heterogeneity, reducing his/her cognitive load when writing queries.
At the same time, we also aim to build a solution that does not excessively increase the query processing time.
We are considering the query components, like literature works, to measure user query complexity and query execution time to measure the database perspective.

%% file: body/sections/c-hk-poly.tex
\section{Hyperknowledge Polystore}
\label{sec:hk-poly}

This section presents our proposal of a polystore architecture that provides users with a single layer for data access encapsulating data and data store heterogeneity, location, and data linkage using schemas metadata, mappings, and provenance.
We named Hyperknowledge Polystore or \hkpoly{} because we use \hk{} (Section~\ref{sec:hyperknowledge}) in its implementation.
Although we could use other ontology representations, \hkpoly{} supports our requirements, like the \textit{Context} composite node we use for knowledge modularization.

%% file: body/sections/d-hkpoly-requirements-and-stakeholders.tex
\subsection{Requirements and Stakeholders}
\label{sec:requirements}

To meet its goal, the Hyperknowledge Polystore supports the following requirements:

\begin{enumerate}[label=(\textbf{R\arabic*}), ref=R\arabic*]
    \item \label{rq:gcs} Create a global conceptual schema (GCS) of the domain; 
    \item \label{rq:gcs-schema} Create a schema representation of the GCS;
    \item \label{rq:lcs} Create local conceptual schemas (LCS) to represent each remote heterogeneous data stores' schemas;
    \item \label{rq:mappings} Create mappings between the GCS and the LCSs;
    \item \label{rq:data-links} Create data links among the data residing in the data stores that are not explicitly linked;
    \item \label{rq:query-processing} Process client queries, \ie{} receive the client query formulated using GCS, create queries considering the LCSs, run the queries on heterogeneous data stores, receive the result, and return to the user.
\end{enumerate}

\hkpoly{} stakeholders and their responsibilities are:
\begin{enumerate}[label=(\textbf{S\arabic*)}]
    \item \textit{Domain Knowledge Engineer}: understands the domain and creates the GCS (\ref{rq:gcs});
    \item \textit{Database Administrators}: create the LCS of each heterogeneous data stores' shared schemas (\ref{rq:lcs});
    \item \textit{\hkpoly{} Knowledge Engineer}: creates a schema for the GCS (\ref{rq:gcs-schema}), and creates the mappings among GCS and LCSs (\ref{rq:mappings});
    \item \textit{Provenance Specialist}: understands the domain processes that manipulate the data residing on the heterogeneous data stores, creates the provenance schema of these processes, and supports developers to instrument the applications that support those processes (\ref{rq:data-links});
    \item \textit{Developers}: instrument the applications that support the domain processes (\ref{rq:data-links});
    \item \textit{Client User}: understands the GCS, formulates queries, and calls \hkpoly{} to process the queries (\ref{rq:query-processing}).
\end{enumerate}

%% file: body/sections/e-hkpoly-architecture.tex
\subsection{Architecture overview}
\label{sec:hkpoly-architecture}

This section presents how \hkpoly{} architecture supports the requirements, depicting the \hkpoly{} metamodel elements.

\hkpolycatalog{} (\hkpoly{} KG) metamodel was created based on W3C-Prov standard and ProvLake (Section~\ref{sec:provenance-provlake}).
We proposed elements inheriting from W3C-Prov (like Collection and Entity), and we use ProvLake elements to represent workflow provenance schema and execution (\eg{} Workflow, Workflow Execution, Data Transformation, Data Transformation Execution, Attribute, Attribute Value).

\begin{figure}[tb]
    \centering
    \includegraphics[trim=0cm 0cm 0cm 0cm, clip=true, width=0.7\textwidth]{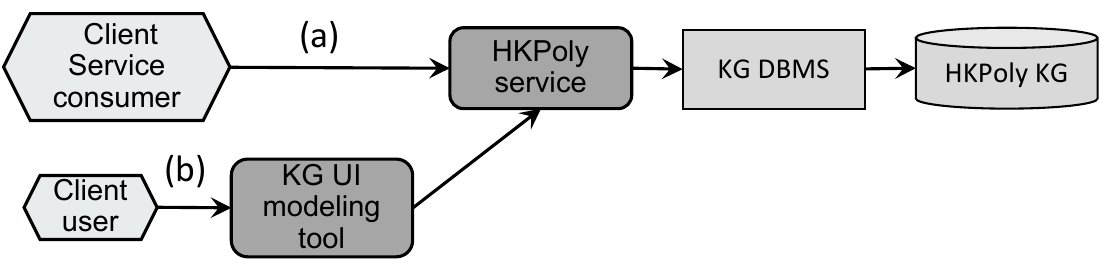}
    \caption{\hkpoly{}-client interaction: (a) Client application calls directly \hkpoly{} service; (b) A user uses a UI which calls the \hkpoly{} service.}
    \label{fig:hk-poly-service-interface}
\end{figure}

\hkpoly{} architecture supports requirements \textbf{\ref{rq:gcs}, \ref{rq:gcs-schema}, \ref{rq:lcs}} and \textbf{\ref{rq:mappings}}  through a service interface for client consumers manage domain and remote data stores metadata (Figure~\ref{fig:hk-poly-service-interface}).
The service is accessed using a client application  (Figure~\ref{fig:hk-poly-service-interface}.a) or by a Web-based UI (Figure~\ref{fig:hk-poly-service-interface}.b).
The received metadata is stored in a database (\hkpoly{} KG), which is managed by Database Management System (KG DBMS). 

\textbf{\ref{rq:gcs} Support:} The \hkpoly{} client user employs a simple language for the GCS creation, \ie{} a language without many constructs.
We propose the use of a Knowledge Graph language (like RDF or Hyperknowledge), using vertices (nodes) to represent concepts and edges to represent relationships among them, respectively (Section~\ref{sec:hyperknowledge}).
As an example, the user represents the concepts \textit{Seismic}, \textit{Horizon} and \textit{Well} as nodes, and the relationships \textit{Seismic has Horizon}, \textit{Seismic has Well} edges among these nodes.
The user may create this model using the \hkpoly{} UI or call \hkpoly{} service.
The GCS is stored in \hkpoly{} KG.

\begin{figure}[tb]
    \centering
    \includegraphics[trim=1.3cm 22.7cm 11cm 1.1cm, clip=true, width=0.4\textwidth]{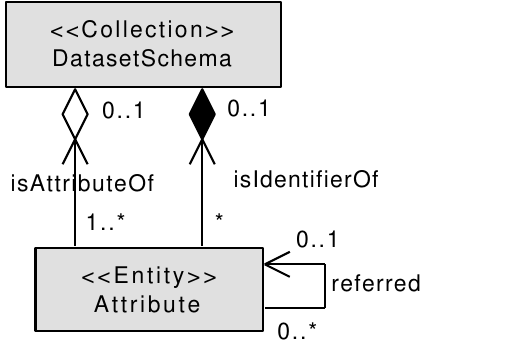}
    \caption{\hkpoly{} model used to represent the GCS schema.}
    \label{fig:hkpoly-gcs-schema}
\end{figure}

\textbf{\ref{rq:gcs-schema} Support:} 
The Knowledge Engineer (KE) formulates a data schema to represent GCS domain elements as instances of the metamodel presented in Figure~\ref{fig:hkpoly-gcs-schema}.
The KE generates \dmelementDataSchema{} for GCS's nodes, \eg{} \textit{Seismic}, \textit{Horizon} and \textit{Well} are created as dataset schemas.
\dmelementAttribute{} represent GCS's edges, \eg{} \textit{hasHorizon} and \textit{hasWell} represent \textit{has Horizon} and \textit{has Well} edges which are related to \textit{Seismic} dataset schema using \dmelement{isAttributeOf} relationship. 
If the attribute identifies the class, the KE also creates the  \dmelement{isIdentifierOf} relationship, \eg{} the attribute \textit{horizonURI} is identifier of \textit{Seismic} dataset schema.
The KE represents the relationship between two concepts using the \dmelement{referred} relationship, \eg{} the attribute \textit{Seismic.hasHorizon} referred the attribute \textit{Horizon.horizonURI}.

\begin{figure}[tb]
    \centering
    \includegraphics[trim=1cm 18.5cm 1.8cm 1.5cm, clip=true,width=0.8\textwidth]{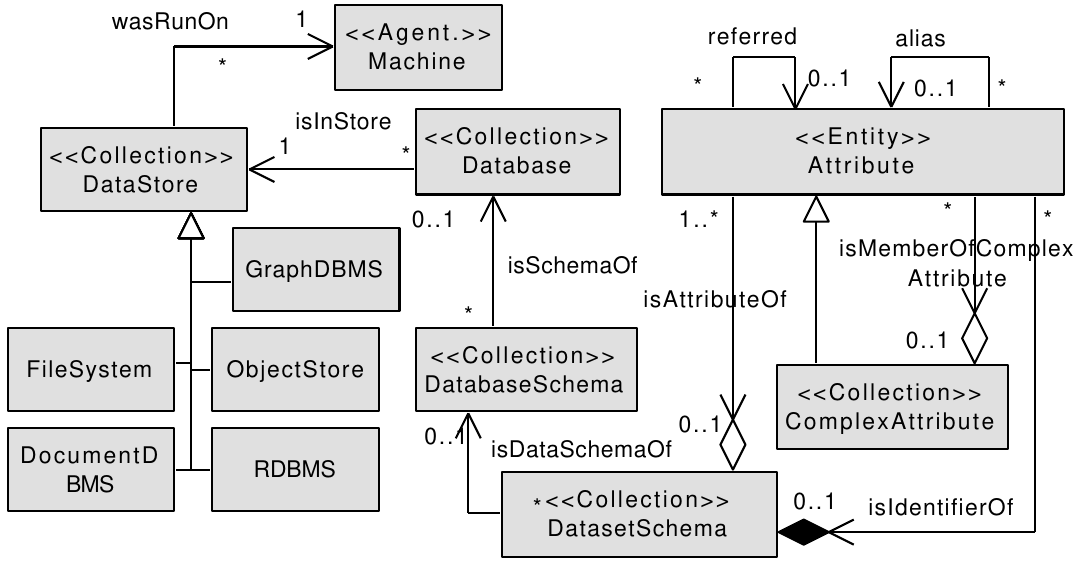}
    \caption{\hkpoly{} model used to represent the LCS schemas.}
    \label{fig:hkpoly-lcs-schema}
\end{figure}

\textbf{\ref{rq:lcs} Support:} The Database Administrator (DBA) creates LCS schemas for each shared portion of the remote data stores' databases using \hkpoly{} template.
This task may be automated by the DBA implementing a script that connects to the data store (\eg{} a MongoDB instance), reads the data store schema, creates the JSON content, and call \hkpoly{} service.
\hkpoly{} service receives the request and creates the LCS schemas as instances of the metamodel presented in Figure~\ref{fig:hkpoly-lcs-schema}, detailed as follows.

A \dmelementDataStore{} may be a \dmelementFileSystem{}, \dmelement{Do\-cu\-ment Da\-ta\-ba\-se}, \dmelement{Re\-la\-tio\-nal Da\-ta\-ba\-se Ma\-na\-ge\-ment Sys\-tem}, \dmelement{Graph Da\-ta\-ba\-se Ma\-na\-ge\-ment Sys\-tem} among others.
A \dmelementDataStore{} runs on a \dmelement{Ma\-chi\-ne} (\dmelement{was\-Run\-On} relationship).

A \dmelementDatabase{} resides in a \dmelementDataStore{} (\dmelement{is\-In\-Sto\-re} relationship),
and it may have \dmelementDatabaseSchema{}s that may have \dmelementDataSchema{}s (\dmelement{is\-Sche\-ma\-Of} and \dmelement{is\-Da\-ta\-Sche\-ma\-Of} relationships, respectively).

\dmelementDataSchema{} and \dmelementAttribute{}s are the same as presented in Figure~\ref{fig:hkpoly-gcs-schema}. This excerpt is used to represent the GCS schema, and also to represent the LCS schema, \eg{} a table and columns of a remote PostgreSQL database.

An \dmelementAttribute{} may be simple or complex (\dmelementComplexAttribute{}) (like a list or a dictionary). 
Simple \dmelementAttribute{}s cannot be subdivided. 
Complex attribute's elements may be composed by other attributes, which composition is represented by \dmelement{is\-Mem\-ber\-Of\-Com\-plex\-At\-tri\-bu\-te} relationship whith \dmelementAttribute{}. 
An \dmelementAttribute{} in a \dmelementSchema{} can be equivalent to another \dmelementAttribute{} in another \dmelementSchema{}, even in different \dmelementDataStore{}s. 
We use \dmelement{ali\-as} to represent equivalent semantics and mappings between attributes, \eg{} an identifier column of a table in a RDBMS may be equivalent to a key of dictionary in a DocumentDBMS.



\textbf{\ref{rq:mappings} Support:} 
The mappings between GCS and LCS elements are performed by the \hkpoly{} KE while creating \dmelement{alias} relationships between GCS and LCS attributes. 
For instance, create an \dmelement{alias}  linking \dmelementeg{Seismic.uri} of GCS with \dmelementeg{Seismic.id} of a relational table of a PostgreSQL LCS.
To accomplish this task, the KE should deeply understand GCS and LCS models. 
In our implementation (presented in Section~\ref{sec:scenario-implementation}),  we have used the \dmelement{Context} construct of \hk{} to modularize the metadata to support the KE navigating in the models and creating the links.

\begin{figure}[tb]
    \centering
    \includegraphics[trim=1cm 17cm 1cm 1cm, clip=true,width=0.8\textwidth]{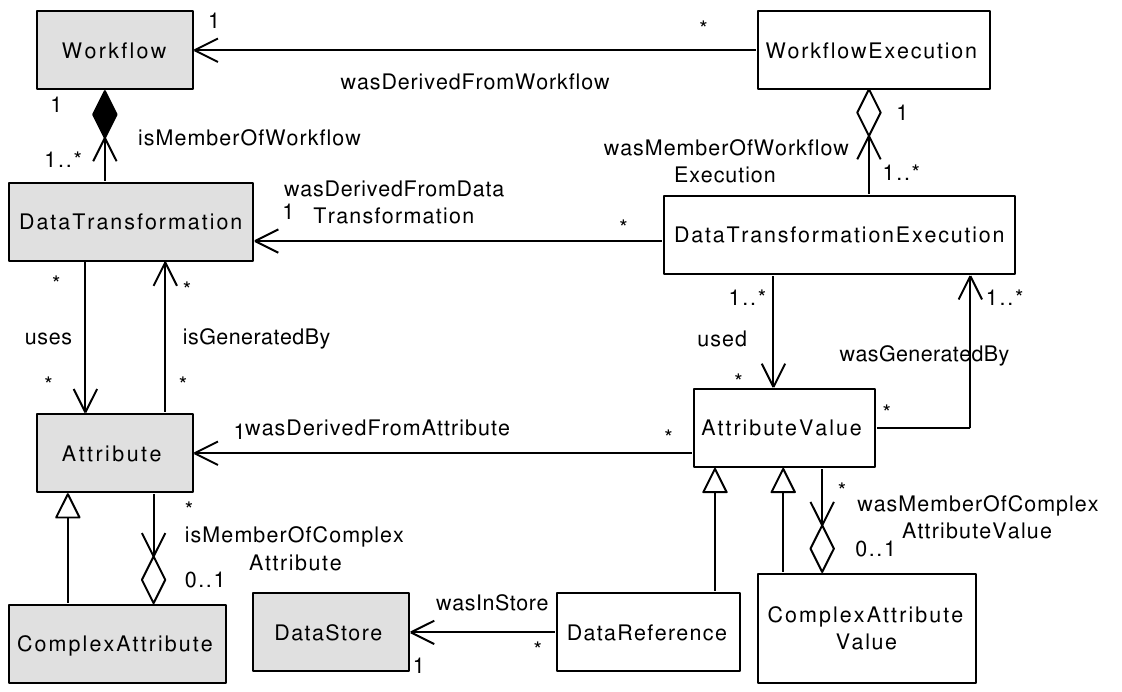}
    \caption{Provenance model based on ProvLake~\cite{souza2019escience}.}
    \label{fig:hkpoly-prov-schema}
\end{figure}

\textbf{\ref{rq:data-links} Support:} 
The linkage between the remote heterogeneous data manipulated by independent workflows is achieved by capturing the provenance of their executions.
Initially, a Provenance Specialist (PS) understands the workflows and creates their provenance schema.
The PS works with Domain Experts and developers of the applications that support the workflows to create the schema\footnote{This model was created inheriting from W3C-Prov as presented by Souza \etal{}~\cite{souza2019escience}. We omitted the stereotypes to simplify the model design.} as instances of the classes presented in gray in Figure~\ref{fig:hkpoly-prov-schema}, such as: \dmelement{Workflow}, its \dmelement{DataTransformation}s and manipulated \dmelementAttribute{}s.

\begin{figure}[tb]
    \centering
    \includegraphics[ width=0.7\textwidth]{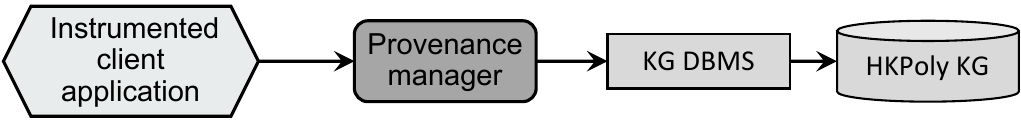}
    \caption{Provenance manager architecture overview.}
    \label{fig:prov-manager-instrumented-workflow}
\end{figure}

Afterwards, the developers include calls to the ProvLakeLib (Section~\ref{sec:provenance-provlake}). in the code to capture the provenance execution and send it to the \dmelement{Provenance Manager} (Figure~\ref{fig:prov-manager-instrumented-workflow}) to store data corresponding to the workflow execution as instances of the classes presented in white in Figure~\ref{fig:hkpoly-prov-schema}, such as: \dmelement{WorkflowExecution}, \dmelement{DataTransformationExecution} and the used and generated \dmelement{AttributeValue}s. 
The \dmelement{DataReference}s are the \dmelement{AttributeValue}s that identify the data records residing in the \dmelement{DataStore}s, \eg{} \dmelementeg{id} and \dmelementeg{URI}.

The schemas metadata, mappings, workflow schema, and workflow execution data stored in \hkpoly{} KG allow the data linkage.
Hence, the mappings among global and local conceptual schemas allow identifying the linkage of instances of different data stores (\ie{} distinct LCSs) representing the same global concept when distinct workflows consume them.
As an example, if one \dmelement{Workflow} has a \dmelement{DataTransformation} that consumes a \dmelementeg{Seismic} file and generates data stored in a PostgreSQL table using \dmelementeg{Seismic.id} as identifier, and it has another \dmelement{DataTransformation} that reads the same file and generates data that is stored in a MongoDB collection using \dmelementeg{Seismic.uri} as identifier, we can infer that these instances represent the same object.

\begin{figure}[tb]
    \centering
    \includegraphics[ width=0.6\textwidth]{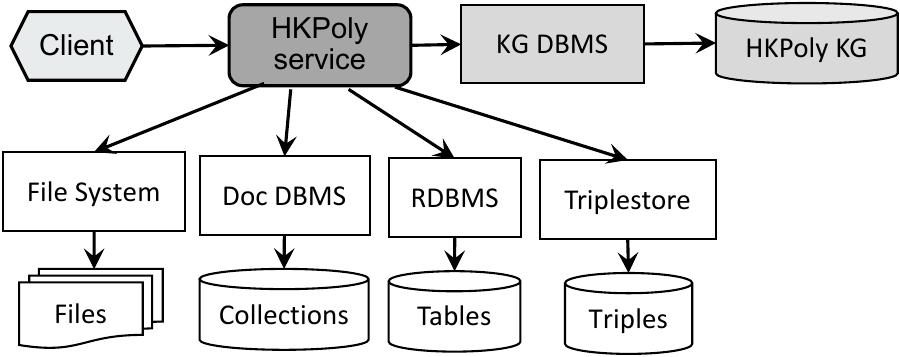}
    \caption{\hkpoly{} architecture overview.}
    \label{fig:hkpoly-query-processing}
\end{figure}

\textbf{\ref{rq:query-processing} Support}: 
An \hkpoly{} architecture overview is depicted in Figure~\ref{fig:hkpoly-query-processing}.
It is based on the Multidatabase Mediator/Wrapper architecture~\cite{ozsu-valduriez:2020:principles-of-data-systems} (Section~\ref{sec:multidatabase}).
To process client-user queries, \hkpoly{} receives a user query formulated considering the GCS concepts and written using an \hkpoly{} supported query language.
Then, \hkpoly{} processes the query as follows:
    (i) It interprets and validates the input query concerning GCS elements, querying the KG, and supported operators;
    (ii) It creates local queries for each LCS that maps to the GCS elements used in the input query - \hkpoly{} queries the KG to get the GCS and LCS mappings and provenance data required for query building;
    (iii) It creates an optimized query execution plan for the local queries;
    (iv) It coordinates the query execution on local DBMSs -  \eg{} File System, Doc DBMS, RDBMS, and Triplestore in Figure~\ref{fig:hkpoly-query-processing};
    (v) It consolidates the results and sends the response to the client-user.

%% file: body/sections/j-hkpoly-implementation-scenario.tex
\section{\hkpoly{} implementation in a real scenario}
\label{sec:implementation}

This section presents the architecture implementation as well as its evaluation in a simulation of a real scenario.
We present the scenario in Section~\ref{sec:scenario},
then \hkpoly{} implementation in Section~\ref{sec:scenario-implementation},
and its evaluation in Section~\ref{sec:scenario-experimental-analysis}.

%% file: body/sections/j.1-scenario.tex
\subsection{Scenario}
\label{sec:scenario}

We evaluate the architecture implementation in the oil reserves discovery, a critical scenario for the Oil \& Gas (O\&G) industry.
We are concerned with seismic image interpretation. 
Typically, these images cover large extensions of the earth, and, by inspecting the images, geoscientists try to identify geological features, \eg{} salt bodies. 
Academia and O\&G industry aim at automating this activity~\cite{randen:2000:seismic-data-analysis}, and deep learning (DL) is a promising ML technique for this~\cite{chevitarese:2018:classification-seismic-textures}.

Managing the data generated during the training of production DL models is hard~\cite{souza:2019:prov-ml}. 
This is particularly true in geoscience problems~\cite{gil:2018:intelligent-system-for-geoscience}. 
It requires preprocessing, cleaning, and performing complex integrated data analysis. 
This lifecycle is decomposed into parts addressed by collaborating teams of geoscientists, computational scientists, engineers, and statisticians, among others. 
Each team has a preferred way to automate tasks and store data, while also having to consume data from other teams, whose preferences might differ.

In particular, the present case study focuses on activities that range from preprocessing large raw geological data files to the generation of training and validation datasets for DL models. 
More details are presented by \cite{souza2019escience,souza:2019:prov-ml}).
In this work, we focus on the heterogeneous data aspect of the use case.
A description of the involved processes is presented in Section~\ref{sec:introduction}.
Figure~\ref{fig:oil-and-gas-wokflow-scenario} illustrates the processes simplified as four chained data processing workflows \cite{souza2019escience}, where each workflow uses one or more data store with heterogeneous data models.

In this case study, the user is an ML expert with deep knowledge in the domain. 
When reporting the results of the ML model, the user must provide domain data about the processed data by the workflows.
An exemplary data to be queried is shown in Table~\ref{tab:query-q}, which illustrates the various data stores that need to be integrated to resolve the query.

\input{body/specific-objects/table-of-queried-data}

%% file: body/specific-objects/table-of-queried-data.tex


\begin{table}[tb]
    \caption{Seismic data and the data stores where they reside.}
    \label{tab:query-q}
    \centering
       \begin{tabular}{@{}cc@{}} 
         \hline \hline
         \textbf{Queried data} & \textbf{Data stores}  \\ [0.5ex] 
         \hline
         Seismic's inline and crossline & PostgreSQL \\ 
         
         Seismic's wells and horizons & AllegroGraph \\
         
          Seismic's geodetic system (EPSG) & Mongo \\
         \hline \hline
        \end{tabular}
\end{table}

%% file: body/sections/j.2-hkpoly-implementation-details.tex
\subsection{ \hkpoly{} Implementation}
\label{sec:scenario-implementation}

This section presents an architecture implementation to support the requirements (Section~\ref{sec:requirements}).
\hkpoly{} data is stored as \hk{} (Section~\ref{sec:hyperknowledge}).
The reason for this choice is to modularize the several kinds of data using the \dmelement{Context} node of \hk{} and reuse the same element in different contexts.
The following tools are used in \hkpoly{} implementation: 
    \hkbase{} and \kes{} (Section~\ref{sec:hkplatform}),  
    ProvLake (Section~\ref{sec:provenance-provlake}), and 
    PostgreSQL Foreign Data Wrapper (FDW) (Section~\ref{sec:postgres-fdw-and-slq-med}).

\hkpoly{} service is implemented as a RESTful web service\footnote{RESTful web services follow REST~\cite{fielding-and-taylor:2002:principled-design-of-modern-architecture} principles to expose cohesive and low coupled web services.} \cite{richardson-and-ruby:2008:restful-web-services}\cite{richardson-etal.:2013:restful-web-apis} provisioned by \hkbase{}. 
This implementation is a realization of the architecture depicted in Figure~\ref{fig:hkpoly-query-processing}.

\begin{figure}[tb]
    \centering
    \includegraphics[trim=1cm 21cm 1cm 1cm, clip=true, width=0.8\textwidth]{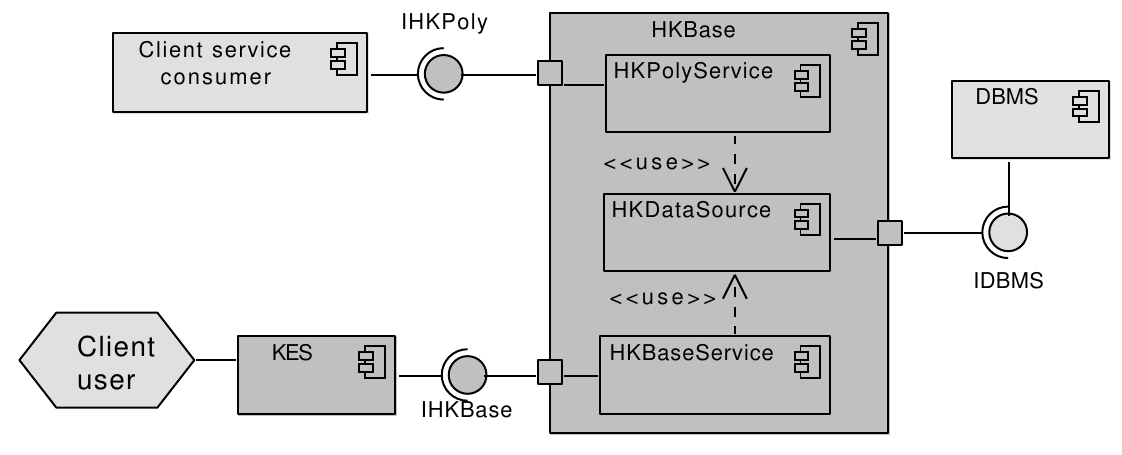}
    \caption{\hkpoly{} service component diagram considering components to support requirements ~\ref{rq:gcs}, \ref{rq:gcs-schema}, \ref{rq:lcs} and \ref{rq:mappings}.}
    \label{fig:hk-poly-ws}
\end{figure}

Figure~\ref{fig:hk-poly-ws} presents a UML component diagram of \hkpoly{} architecture to support \textbf{\ref{rq:gcs}, \ref{rq:gcs-schema}, \ref{rq:lcs},} and \textbf{\ref{rq:mappings}} through two mechanisms: Service Interface, and UI tool.
\hkpoly{} Service Interface is provisioned as part of \hkbase{} API, \ie{} \hkpoly{} endpoints are provisioned as part of \hkbase{} contract - \dmelement{IHKPoly} interface in the diagram.
Client applications call \hkpoly{} service to create domain model, GCS, LCS, and mappings among them.
\hkpolyservice{} uses \hkbase{}'s \dmelement{HKDataSource} to store \hkpolycatalog{} data in its main DBMS. 
The main DBMS being the one supported by the \hkbase{} instance being used, \eg{} Apache Jena\footnote{\url{https://jena.apache.org/}} RDF store.
In the UI mechanism, a client may use the \kes{} tool to visualize and manipulate the ingested data.

The domain model may be created using basic elements of \hk{} (like \dmelement{Concept node} and \dmelement{Connectors} for relationships), and the GCS, LCS, and mappings are created using the models presented in Section~\ref{sec:hkpoly-architecture} in Figure~\ref{fig:hkpoly-gcs-schema} and Figure~\ref{fig:hkpoly-lcs-schema}.

As an example, consider the domain model illustrated in Table~\ref{tab:query-q}, \ie{}  \dmelementeg{Seismic} node and edges connecting it with \dmelementeg{name}, \dmelementeg{inline}, \dmelementeg{crossline}, \dmelementeg{well}, \dmelementeg{horizon}, and \dmelementeg{epsg} properties.
\hkpoly{} creates a GCS for the domain using the proposed schemas, which data is illustrated in Table~\ref{tab:gcs-and-lcs} first line (after the header).

Consider also we have \dmelementeg{Seismic} data stored in the following remote heterogeneous data stores, as follows:
\begin{itemize}
    \item PostgreSQL: \dmelementeg{SeismicHeader} table with columns \dmelementeg{id}, \dmelementeg{inline}, \dmelementeg{crossline}, \dmelementeg{header\_info}, \dmelementeg{filepath};
    \item AllegroGraph: \dmelementeg{SeismicCls} class with properties \dmelementeg{URI}, \dmelementeg{name}, \dmelementeg{hasWell}, and \dmelementeg{hasHorizon};
    \item MongoDB: \dmelementeg{Seismc\_data} collection with fields \dmelementeg{identifier}, \dmelementeg{name}, \dmelementeg{num\_ilines}, \dmelementeg{num\_xlines}, \dmelementeg{epsg}.
\end{itemize}
Table~\ref{tab:gcs-and-lcs} (lines 3, 4, and 5) presents the LCS remote data store schemas, and Table~\ref{tab:gcs-and-lcs-mappings} presents the mappings between GCS and LCS. 
These mappings are done using \dmelement{alias} relationship of \hkpoly{} model (Figure~\ref{fig:hkpoly-lcs-schema}).

\input{body/specific-objects/table-of-gcs-and-lcs}

\textbf{\ref{rq:data-links}} is supported by provenance manager component, named as \hkpolyprovmanager{}, provisioned also as a RESTful web service.
The implementation is a realization of the overview architecture presented in Figure~\ref{fig:prov-manager-instrumented-workflow}.

\begin{figure}[tb]
    \centering
    \includegraphics[trim=1cm 22.5cm 2cm 1cm, clip=true, width=0.8\textwidth]{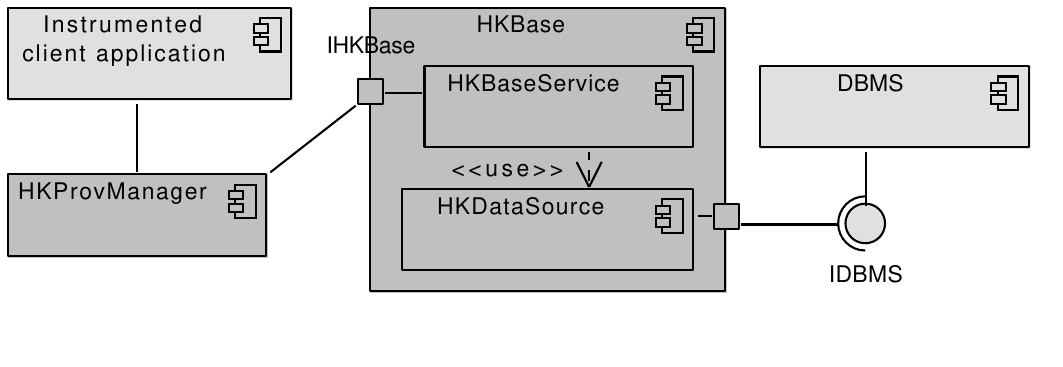}
    \caption{Provenance Manager component architecture.}
    \label{fig:hkprov-manager}
\end{figure}

Figure~\ref{fig:hkprov-manager} presents a component diagram of \hkpolyprovmanager{}.
So, the developers of client applications create hooks in their code (illustrated by the \dmelement{Instrumented client application} component of Figure~\ref{fig:hkprov-manager}) which make calls to \hkpolyprovmanager{} service.
\hkpolyprovmanager{} uses \hkbase{} services to store the provenance data, using the \dmelement{HKDataSource} to access the DBMS that stores \hkpolycatalog{}.
The service calls aim to store provenance execution data, like:
\begin{enumerate}
    \item \dmelement{Workflow} start information;
    \item \dmelement{Workflow} \dmelement{DataTransformation} executions and their input parameters (\dmelement{used Attributes and AttributeValues}) and returned data (\dmelement{generated Attributes and AttributeValues})
    \item \dmelement{Workflow} end information.
\end{enumerate}

The \dmelement{Workflow} schema and its counterparts are created when execution data is loaded, or a \dmelementeg{Provenance Expert} may use the \dmelement{KES} to create this schema.
So, the only requirement of the instrumented code is that the \dmelement{Attributes}' names are the same as those defined in the LCS.
It allows the identification of which domain data were consumed by each workflow and, consequently, the linkage among the object fragments residing in each data store.
These data are retrieved when \hkpoly{} handles client queries, and it is combined with remote data stores' data to answer the user request. 
Details are presented when explaining how \hkpoly{} supports Requirement~\ref{rq:query-processing}.

\begin{figure}[tb]
    \centering
    \includegraphics[trim=1cm 21cm 1cm 1cm, clip=true, width=0.9\textwidth]{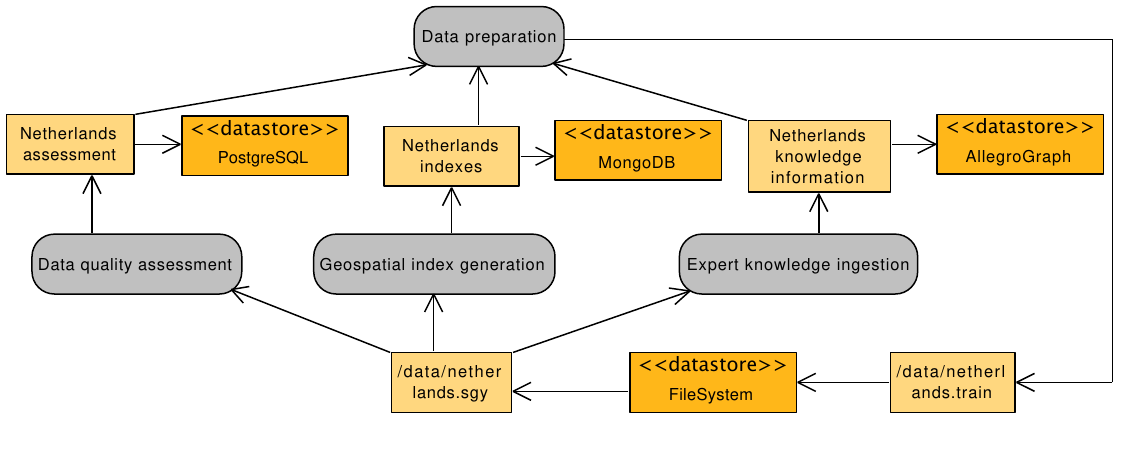}
    \caption{Example of execution of the workflows of Figure~\ref{fig:oil-and-gas-wokflow-scenario} to investigate the Seismic Netherlands. Workflows are illustrated as actions in the diagram.}
    \label{fig:workflow-execution}
\end{figure}

Figure~\ref{fig:workflow-execution} presents a UML Activity diagram to exemplify the execution of the workflows of Figure~\ref{fig:oil-and-gas-wokflow-scenario} to process the Seismic \textit{Netherlands}.
Workflows are presented as actions for simplification.
The first workflow \dmelementeg{Data quality assessment} reads the \dmelementeg{netherlands segy} file from the \dmelementeg{FileSystem} and inserts \dmelementeg{Netherlands assessments} in a \dmelementeg{PostgreSQL} database table.
The second workflow \dmelementeg{Geospatial index generation} reads the same file and stores \dmelementeg{Netherlands indexes} in a \dmelementeg{MongoDB} collection.
The third workflow \dmelementeg{Expert knowledge ingestion} stores \dmelementeg{Netherlands knowledge information} in AllegroGraph.
An Expert provides this information when analyzing the \dmelementeg{netherlands segy} seismic file.
Finally, the fourth workflow \dmelement{Data preparation} creates the training data by reading all the stored data and generating the \dmelementeg{netherlands.train} file which is stored in the \dmelementeg{FileSystem}.
During this workflow execution, while storing data in the data stores, the instrumented client application also sends provenance data calling \hkpolyprovmanager{}, which stores the \dmelement{DataReferences} for the record, document, and triples (id, identifier, and respectively) in \hkpolycatalog{}.
Table~\ref{tab:workflow-execution-data} presents the \dmelement{DataReferences} captured during the execution of Netherlands workflow.
The provenance data, GCS, LCSs, and mappings are used to compute the polystore query detailed in \textbf{\ref{rq:query-processing}} support implementation.

\input{body/specific-objects/table-of-workflow-execution-data}

\begin{figure}[tb]
    \centering
    \includegraphics[trim=1cm 19cm 1cm 1cm, clip=true, width=0.8\textwidth]{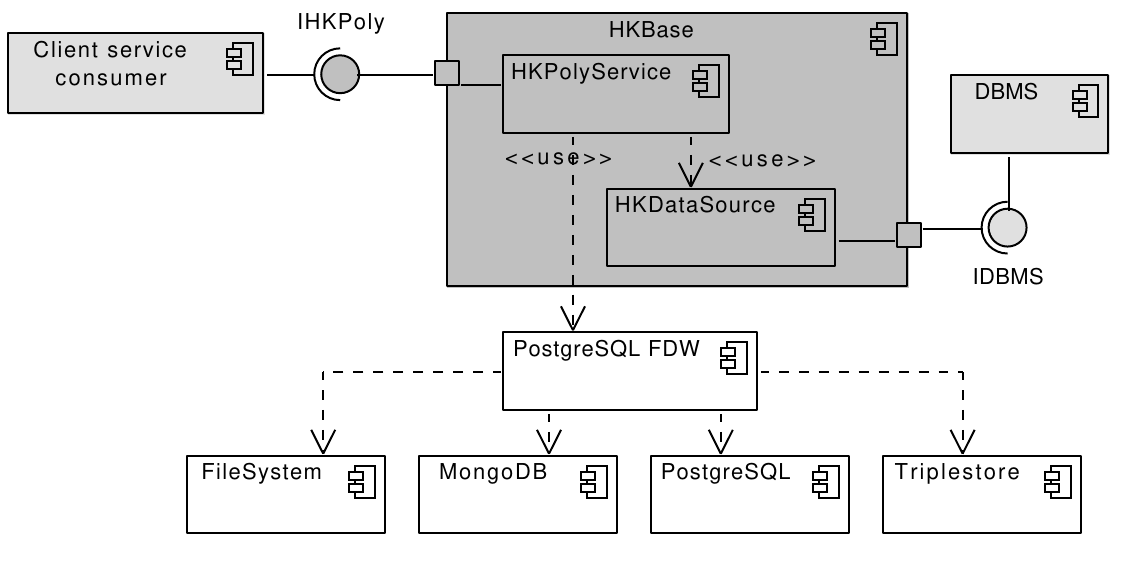}
    \caption{ \hkpoly{} component diagram considering the \ref{rq:query-processing} support.}
    \label{fig:hkpoly-query-execution}
\end{figure}

Figure~\ref{fig:hkpoly-query-execution} presents \hkpoly{} component diagram concerning the support for \textbf{\ref{rq:query-processing}}.
It is a realization of the architecture depicted in Figure~\ref{fig:hkpoly-query-processing}.
The client application creates a query based on the GCS model using the Hyperknowledge query language (\hyql{}\footnote{The \hyql{} grammar is presented in \url{https://ibm.ent.box.com/v/iswc2021-hyql-grammar}}) \cite{soares-et-al:2021:hk-dataset-engineering} (Section~\ref{sec:hyperknowledge}).
An example is Query~\ref{lst:hyql-scenario} which retrieves \dmelementeg{Seismic} attributes (lines 1 and 2) considering workflow \dmelementeg{geological\_data\_ingestion\_workflow} (Line 3) and \dmelementeg{Seismic}'s \dmelementeg{name} is \textit{Netherlands} (Line 4).
The workflow is referenced using the \dmelement{from} clause, which indicates that queried data is contained in a \dmelement{Context} object.

\input{body/specific-objects/query-hyql}

\hkpoly{} interprets and validates the query using the data stored in the \hkpolycatalog{}.
Query~\ref{lst:sparql-to-get-wf-and-gcs-data} is a example of \hkpoly{} service query used to retrieve \dmelement{KG}'s data.
It retrieves workflow execution and related GCS data (lines 1, 2, and 3), besides Foreign Data Wrapper data (lines 4 and 5), which is used by \hkpoly{} to create a polystore query which will be explained afterward.
The SPARQL language is used to navigate in \dmelement{KG} executing, \eg{} query path and inverse property navigation (lines 10 and 11) and traversing the workflow elements (lines 8 to 15) and GCS (lines 16 to 19). 
The \dmelement{att} variable is used to connect workflow and GCS data (Line 10 for workflow and Line 16 for the GCS).

To access the remote heterogeneous data stores, \hkpoly{} uses \dmelement{PostgreSQL Foreign Data Wrapper} (FDW) (Section~\ref{sec:postgres-fdw-and-slq-med}).
\hkpoly{} does not implement a new polystore connector interface since there are some available, and the work focus on validating the use of the GCS and provenance to enable a seamless access interface to users regarding abstraction and data linkage. 
So, we use \dmelement{PostgreSQL FDW} as our polystore connector since it has a broad use and several foreign data wrappers.
Hence, the \dmelement{PostgreSQL FDW} schemas are ingested as LCS, \ie{} the remote data stores' shared data are represented as LCS of \dmelement{PostgreSQL FDW data}, and GCS is liked to these LCSs.
The links are created using \dmelement{alias} relationship which is used in Query~\ref{lst:sparql-to-get-wf-and-gcs-data} (Line 20) to reach FDW data in lines 21 to 24.
After getting this information, \hkpoly{} computes the SQL presented in Query~\ref{lst:polystore-sql}.
This query is sent to \dmelement{PostgreSQL FDW} which process the results and returns to \hkpoly{} the response to the \dmelement{Client service consumer}.

\input{body/specific-objects/query-sparql-catalog}

In Query~\ref{lst:polystore-sql}, the \dmelement{from} clause (lines 6 to 8) lists the \dmelement{PostgreSQL Foreign Tables}\footnote{Foreign tables behave like regular tables, and they are used to create mappings in a PostgreSQL database to external objects~\cite{melton:2001:sql-med}.} that maps to the remote data stored in a \dmelementeg{segy file}, \dmelementeg{AllegroGraph knowledge base triples}, \dmelementeg{Mongo Seismic collection}, and \dmelementeg{Seismic\_Header table}.
Lines 9 to 13 compute a ``\textit{constant table}'' using the \dmelement{VALUES}\footnote{\url{https://www.postgresql.org/docs/current/sql-values.html}} SQL clause with the workflow executions data values retrieved using Query~\ref{lst:sparql-to-get-wf-and-gcs-data}. 
In the \dmelement{where} clause, the \dmelementeg{Foreign Tables} and \dmelementeg{constant table} are joined (lines 14 to 17), and, as \dmelementeg{Seismic.name} is linked to \dmelementeg{kb\_seismic} and \dmelementeg{seismic\_header} \dmelementeg{Foreign Tables}, lines 18 and 19 are included in the SQL during its generation.
The returned data is present in lines 1 to 5, which references the columns of the \dmelementeg{Foreign Tables} mapping to \dmelementeg{Seismic} domain object.

\input{body/specific-objects/query-polystore}


%% file: body/specific-objects/table-of-gcs-and-lcs.tex
\begin{table}[tb]
    \caption{GCS of Seismic (line 1) and LCS of the Seismic data residing in the remote stores (lines 2, 3, and 4).}
    \label{tab:gcs-and-lcs}
    \centering
\begin{tabular}{@{}llllcp{0.18\textwidth}c@{}}
\toprule
Line & DataStore & Database & Schema & DatasetSchema & \multicolumn{1}{c}{Attribute}             & \multicolumn{1}{c}{Identifier} \\
~ & ~ & ~ & ~ & ~ & \multicolumn{1}{c}{(isAttributeOf)}             & \multicolumn{1}{c}{(isIdentifierOf)} \\
\midrule
1 & ~ & ~ & ~ & Seismic & URI, inline, crossline, well, horizon, epsg & URI         
\\
2 & PostgreSQL  & SeismicDB &  SeismicSQ  & SeismicHeader &  id, inline, crossline, header\_info, filepath &  id 
\\
3 & AllegroGraph & Seismic catalog & Seismic repo &  SeismicCls & URI, name, hasWell, hasHorizon & URI 
\\
4 & MongoDB & Seismicdb & Seismic & Seismic\_data & identifier, name, num\_ilines, num\_xlines, epsg & identifier
\\
\bottomrule
\end{tabular}
\end{table}

\begin{table}[tb]
    \caption{Seismic GCS and LCS mappings attribute mappings.}
    \label{tab:gcs-and-lcs-mappings}
    \centering
       \begin{tabular}{@{}cc@{}} 
         \hline \hline
         \textbf{GCS} & \textbf{LCS}  \\
         \hline
            Seismic.URI & SeismicHeader.id \\
         
            Seismic.URI & SeismicCls.id \\
         
            Seismic.URI & Seismic\_data.identifier \\
         
            Seismic.inline & SeismicHeader.inline \\
         
            Seismic.crossline & SeismicHeader.crossline \\
         
            Seismic.well & SeismicCls.hasWell \\
         
            Seismic.horizon & SeismicCls.hasHorizon \\
         
            Seismic.epsg & Seismic\_data.epsg \\
         \hline \hline
        \end{tabular}
\end{table}

%% file: body/specific-objects/table-of-workflow-execution-data.tex
\begin{table*}[tb]
    \caption{Remote data store data captured during Seismic interpretation DL workflow execution.}
    \label{tab:workflow-execution-data}
    \centering
       \begin{tabular}{@{}cccc@{}} 
         \hline \hline
         \textbf{\dmelement{DataTransformation}} & \textbf{\dmelement{DatasetSchema}} & \textbf{\dmelement{Attribute}} & \textbf{\dmelement{AttributeValue}}  \\
         \hline
            Data quality assessment & SeismicHeader & id & 12345\\
         
            Geospatial index generation & Seismic\_data & identifier & 1111 \\
         
            Expert Knowledge Ingestion &  SeismicCls & URI & \url{http://oilandgas/Seismic#Netherlands} \\
        
            Data preparation & Training File & path & \url{/data/netherlands.train} \\
         \hline \hline
        \end{tabular}
\end{table*}

%% file: body/specific-objects/query-hyql.tex
\begin{lstlisting}[basicstyle=\ttfamily\scriptsize, breaklines=true, numbers=left, xleftmargin=0.5cm, escapechar=-,caption={HyQL query to retrieve Netherlands seismic data.}, float, label=lst:hyql-scenario]
select Seismic.inline, Seismic.crossline, 
Seismic.hasWell, Seismic.hasHorizon, Seismic.epsg
where Seismic from geological_data_ingestion_workflow 
and Seismic.name = "Netherlands"
\end{lstlisting}




%% file: body/specific-objects/query-sparql-catalog.tex
\begin{lstlisting}[basicstyle=\ttfamily\scriptsize, breaklines=true, numbers=left, xleftmargin=0.5cm, escapechar=-, caption={SPARQL query to retrieve workflow execution and LCS data.}, float, label=lst:sparql-to-get-wf-and-gcs-data]
select  distinct ?wfe ?atv ?atvValue ?att ?attName 
  ?datasetSchema ?datasetSchemaName 
  ?dataStore ?dataStoreName
  ?attFDW ?attFDWName 
  ?datasetSchemaFDW ?datasetSchemaFDWName
where
{
  ?atv <hk://id/wasDerivedFromAttribute> ?att.
  ?att <hk://id/name> ?attName.
  ?atv <hk://id/wasGeneratedBy>
    |^<hk://id/used> ?dte.
  ?atv <hk://id/value> ?atvValue.
  ?dte <hk://id/wasMemberOfWorkflowExecution> ?wfe.
  ?wfe <hk://id/wasDerivedFromWorkflow> 
      <hk://id/${workflow}>.
  ?att <hk://id/isAttributeOf> ?datasetSchema.
  ?datasetSchema <hk://id/name> ?datasetSchemaName.
  ?att <hk://id/isStoredInStore> ?dataStore.
  ?dataStore <hk://id/name> ?dataStoreName.
  ?attFDW <hk://id/alias> ?att.
  ?attFDW <hk://id/name> ?attFDWName.
  ?attFDW <hk://id/isAttributeOf> ?datasetSchemaFDW.
  ?datasetSchemaFDW <hk://id/name> 
    ?datasetSchemaFDWName.
}
order by ?wfe ?dte ?atv
\end{lstlisting}

%% file: body/specific-objects/query-polystore.tex
\begin{lstlisting}[basicstyle=\ttfamily\scriptsize, breaklines=true, numbers=left, xleftmargin=0.5cm, escapechar=-, caption={SQL to get data of Netherlands seismic.}, float, label=lst:polystore-sql]
SELECT distinct fdw_kb_seismic."hasWell", 
	fdw_mongo_seismic."epsg", 
	fdw_seismic_header."crossline", 
	fdw_kb_seismic."hasHorizon", 
	fdw_seismic_header."inline" 
FROM segy fdw_segy, kb_seismic fdw_kb_seismic, 
	mongo_seismic fdw_mongo_seismic, 
	seismic_header fdw_seismic_header,
	( VALUES ( 'netherlands.sgy', 
	'http://br.ibm.com/hkpoly/seismicData_ABox#Netherland_3D', 	
	'http://br.ibm.com/hkpoly/seismicData_ABox#Netherland_3D', 1 )) 
	as p(LocalFileSystem1_prov_id, Allegro1_prov_id, 
	Mongo1_prov_id, Postgres1_prov_id)
WHERE fdw_kb_seismic.uri=p.Allegro1_prov_id 
	AND fdw_mongo_seismic.uri=p.Mongo1_prov_id 
	AND fdw_seismic_header.seismic_header_id=
	    p.Postgres1_prov_id
	AND fdw_kb_seismic.name = 'Netherlands'
	AND fdw_seismic_header.name = 'Netherlands'
\end{lstlisting}



%% file: body/sections/j.4-evaluation-considering-this-scenario.tex
\subsection{Evaluation}
\label{sec:scenario-experimental-analysis}



Our evaluation measures query complexity considering the two perspectives described in Section~\ref{sec:query-complexity} : the user perspective and the database perspective.

\subsubsection{Query Complexity - User's Perspective}

We choose counting the query components to estimate the user's cognitive load measurement.
It is an approach presented in some works as showed in Section~\ref{sec:query-complexity}.
We do not choose a weighted sum because there is no convergence that a weighting technique would improve the evaluation. 
We infer that counting the query components, regardless of the summing technique, is a fair estimate to evaluate the user's cognitive load.

In this case, we count query components as an estimate to measure the user's cognitive load while writing a query.
As described in Table~\ref{tab:cognitive-load}, the user query (Query~\ref{lst:hyql-scenario}), written in \hyql{} language, has 5 elements in the projection, 1 element in \texttt{from} clause, and 1 query filter. 
It results in 7 query components.
On the other hand, the equivalent polystore SQL query (Query~\ref{lst:polystore-sql}) has 5 elements in the projection, 5 elements in \texttt{from} clause, 3 \texttt{joins} and 1 query filter, resulting on
14 query components \footnote{We are considering the constant table of Listing~\ref{lst:hyql-scenario}, created using \texttt{Values} clause, counting as 1 since its data can be encapsulated, \eg{} in a temporary table, making the query easier to formulate.}.
Hence, in this example, we can infer that the query written in \hyql{} is less complex than the SQL query generated by \hkpoly{} to access the external data.

\input{body/specific-objects/table-cognitive-load}

\subsubsection{Query Complexity - Database's Perspective}
In this case, we evaluate query complexity by measuring its processing time.
Therefore, we evaluate the processing time \hkpoly{} takes to process the user query, \ie{} 
the implementation depicted in Figure~\ref{fig:hkpoly-query-execution} to support \textbf{\ref{rq:query-processing}}.
We compared this time against the time required to process solely the \postgresqlfdw{} query.
The \hkpoly{} processing time includes:
    (i) receive the query (through \ihkpoly{} interface), \eg{} Query~\ref{lst:hyql-scenario};
    (ii) parser the received query;
    (iii) run queries over \hkpolycatalog{} (using \hkdatasource{}) to get required metadata and mappings, running Query~\ref{lst:sparql-to-get-wf-and-gcs-data} and others;
    (iv) compute the \postgresqlfdw{} query, \eg{} Query~\ref{lst:polystore-sql}; and,
    (v) \label{step:postgresqlfdw-processing} run the computed query using \postgresqlfdw{}.
The compared \postgresqlfdw{} time corresponds  only  to  the time elapsed in Step~\ref{step:postgresqlfdw-processing}.
We vary the data volume of \hkpolycatalog{} and the data volume of remote data stores to analyze \hkpoly{}  overhead when remote data stores' data volume increases.

The first step to verify our hypothesis is to build an experimental deployment of our envisioned \hkpoly{} architecture implementation. 
For that, we use Docker\footnote{\url{https://www.docker.com/}} container solution, allocating 5 CPUs, 10GB RAM, and a swap area of 2GB in a machine with i5 Intel Quad Core 2GHz CPU and 16 GB RAM.
\hkbase{}, and its services (including \hkpolycomponent{\hkpoly{}} and \hkdatasource{}) are deployed in a container.
Our test bed mirrored the proposed architecture presented in Figure~\ref{fig:hkpoly-query-execution}, considering that all servers were part of the same network. 
In this setup, we use \hkpolycomponent{Apache Jena} as \hkbase{}'s main DBMS, and it is deployed in a single container. 
Each remote data store is also deployed in a separate container using a proper schema to match the seismic domain model, corresponding to the scenario presented in Section~\ref{sec:scenario} and used to describe \hkpoly{} implementation (Section~\ref{sec:scenario-implementation}). 
In this setup, we have the seismic data residing on \hkpolycomponent{AllegroGraph}, \hkpolycomponent{MongoDB}, and \hkpolycomponent{PostgreSQL}.
Table~\ref{tab:gcs-and-lcs} exemplifies Seismic attributes of data on each data store.
The \postgresqlfdw{} is deployed in a single container and loaded with data wrappers plugins to provide communication and data transformation between the remote databases. 
All these containers are started altogether with a composing script.

After the test environment is set up, we load domain data into the remote stores and provenance data into \hkpolycatalog{}.
To control the volume of data, we generate domain data in what we call a batch. 
A batch has 16KB and represents three different files generated by a synthetic data generator whose schemas match the remote databases' schemas.
They correspond to the data used by the three data transformations presented in Figure~\ref{fig:workflow-execution}.
After being generated, we simulate the data transformations' execution and load the results in \hkpolycomponent{PostgreSQL}, \hkpolycomponent{MongoDB}, and \hkpolycomponent{AllegroGraph}, respectively.
The generated provenance is loaded into \hkpolycatalog{}.
Afterward, we start our performance tests. 

We choose to run the test of query processing time performance using different numbers of batches varying the data volume to understand its influence on the processing. 
We start with one batch (B001) and then increase to 50, 100, 400, and 700. 
For each query performance test, we purge the provenance execution data from the \hkpolycatalog{} repository, keeping the schemas (GCS, LCS, provenance, and mappings schemas), and erase the remote databases' contents.
Also, for each batch quantity, the query presented in Query~\ref{lst:hyql-scenario} is executed fifty times, so we can calculate the aggregated statistics for each one, waiting for one second between queries requests.  
According to Hoefler and Belli \cite{hoefler-et.al.:2015:scibmk}, the academic literature suggests that sample sizes between thirty and forty are sufficient for normal distributions. 
Although we have not verified that the query processing time is normally distributed, we use the sample size of fifty based on this assumption.

\begin{figure}[tb]
    \centering
    \includegraphics[width=0.8\textwidth]{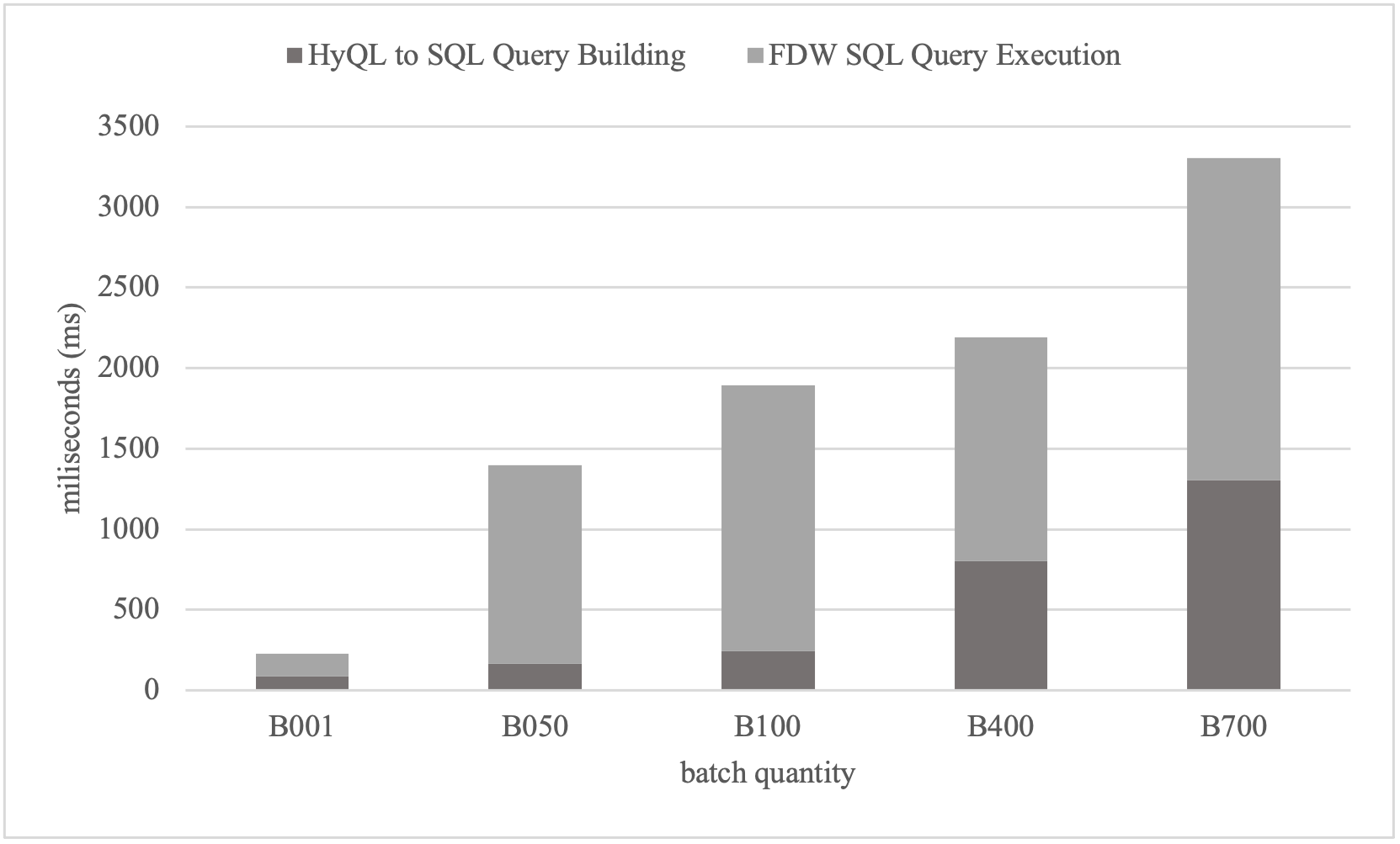}
    \caption{HyQL to SQL Query Building and FDW SQL Query Execution Response Time Analysis Over Batch Quantities (ms).}
    \label{fig:expResult02}
\end{figure}

Finally, we instrument \hkpoly{} service to capture query execution times to understand the impact of building the query to be executed on the \postgresqlfdw{} component (which we name as \buildingquery{} step), \ie{} transform the input query (\eg{} Query~\ref{lst:hyql-scenario} to Query~\ref{lst:polystore-sql} in the case of one batch). 
We collect the \buildingquery{} step time and the time to run the query in \postgresqlfdw{}, whose sum corresponds to the whole query execution time. 
We assume that the time of the \hkpolyclient{} to send the query to \hkpoly{} and the time of \hkpoly{} to send the response back is negligible.
The result of the query response time experiments for each batch's quantities is shown in Figure~\ref{fig:expResult02}. It presents the medians of the overall query response time extracted from the fifty rounds executed on each batch experiment. 
It shows this response time split into two parts: one for the query building (\buildingquery{}) and another one for the \postgresqlfdw{} query execution (\fdwqueryexecution{}). 

Although it is straightforward to assume that the overall query response time would increase due to the increasing loaded data, the time consumed to build the \postgresqlfdw{} SQL query is not the ruling part of the process. 
In the worst case, our experiments show that the transformation process takes less than 30\% of the overall query processing time.
We expect that this behavior remains the same whenever more data is loaded into the \hkpoly{} environment.

%% file: body/specific-objects/table-cognitive-load.tex
\begin{table}[tb]
    \caption{Accounting query components as an estimate of user's cognitive load.}
    \label{tab:cognitive-load}
    \centering
        \begin{tabular}{@{}l c c@{}}
        \hline \hline
        \multicolumn{1}{c}{Query Components} & Query 1    & Query 3     \\ 
        \hline
Projection                             & 5          & 5            \\ 
Filter                                 & 1          & 1            \\ 
Join Clause                            & 0          & 3            \\ 
From Clause                            & 1          & 5            \\ 
        \textbf{Total}                         & \textbf{7} & \textbf{14} \\ 
        \hline \hline
        \end{tabular}
\end{table}

%% file: body/sections/n-related-work.tex
\section{Related Work}
\label{sec:related-work}

Several works have tackled the database federation problem~\cite{azevedo-et-al:2020:modern-federated-ds-overview}~\cite{tan:2017}. 
The Garlic system~\cite{carey::1995::towards-garlic} is capable of integrating data from a broad range of data repositories. 
Garlic's architecture is based on repository wrappers, an object-oriented data model and query language to provide a uniform view of heterogeneous data types and data sources.
The DiscoveryLink system~\cite{haas::2001::discovery-link} allows users to query data stored in heterogeneous and physically distributed data stores by using a virtual database. 
DiscoveryLink system is based on the fusion of several DB2~\cite{haas::2002::data-integration-db2} and Garlic~\cite{carey::1995::towards-garlic} components. 

While Garlic and DiscoveryLink use SQL, \cite{langegger::2008::semantic-web-middleware} propose a middleware that employs SPARQL, 
 shares architectural principles with Garlic's architecture, and enables integration of CSV files and Relational Databases with RDF data sources. 
This integration is achieved by the implementation of RDF wrappers that are very similar to the repository wrappers of Garlic and DiscoveryLink.
\cite{le::2012::middleware-linked-streams} proposal uses SPARQL and integrates time-dependent data with other RDF data sources by enriching both sensor sources and sensor data streams with semantic descriptions.

The Information Integrator tool~\cite{angele:2006:data-integration}  uses ontologies and F-Logic to create an integrated view of data and mapping between data objects in distributed heterogeneous data sources, \eg{} databases, web services, and applications. 
The solution is divided in four layers:
    (i) Data sources;
    (ii) Ontologies that represent each data source;
    (iii) Business ontology that provide a conceptualization of business entities. 
    (iv) views (\ie{} queries) of the business ontology. 
Elements of the layers are connected by mappings.
F-Logic rules are used to specify mappings that assemble higher-value business ontologies from others. 

\hkpoly{} has similarities with these proposals, but it aims for simplification, expressiveness gain, and connecting heterogeneous data that do not have explicit links using provenance.
The use of provenance to link data manipulated by independent workflows is not considered in any previous works.
Using our proposal, the user specifies a query considering a single model and indicates from which workflow (or workflow execution) to get the data. 
Then, \hkpoly{} discovers the paths to navigate and the data sources to reach.

Several years after those initial attempts to address the problem of heterogeneous data federation, the BigDAWG polystore system~\cite{gadepally-et-al:2016:bigdawg} handles the competing notions of location transparency and semantic completeness when supporting diverse database systems working with different data models. 
For this purpose, the authors leveraged the concepts of islands and shims to create a middleware that provides a uniform query interface while allowing users to exploit the full capabilities of each database connected to the system. 
One can argue that BigDAWG still requires the implementation of so-called degenerate islands to support the full semantic power of a connected database, which is somewhat similar to the repository wrapper-based architecture of previous solutions. 
As novelty, this system does not require the user to commit to one specific data model, such as Relational or RDF, as its base query language mainly comprises two general operators (CAST and SCOPE) that can be used to combine queries following multiple data models.

Although BigDAWG represents a breakthrough and has been evaluated with success in multiple medical applications, many large-scale applications continued to use federated database systems that extended the concept of repository wrappers used in Garlic and DiscoveryLink and used SQL as their primary query language. 
One of the most popular examples is PostgreSQL, which supports the implementation of Foreign Data Wrappers (FDWs) for external data sources access. 
These FDWs are mainly responsible for converting the data stored in the heterogeneous data sources into a relational representation, while the PostgreSQL engine is responsible for processing the multiple joins required to answer a query that integrates their corresponding tables. 
Nonetheless, this system allows the implementation of condition pushdowns in which conditions clauses of SQL queries, described after the WHERE clause, are directly translated to operations in the native query language of each data source, enabling significant gains in query performance~\cite{wang::2017::postgressql-fdw}.

BigDAWG and PostgreSQL represent state-of-the-art solutions to the problem of database federation -
BigDAWG mainly as a research prototype and PostgreSQL FDW as an industry solution.
However, we consider that they can still not address the main requirements posed by modern applications since BigDAWG requires the user to understand the query languages of the multiple federated databases integrated through the system. Also, it requires how to combine them using its basic operators, while PostgreSQL requires the user to know the schema of the FDW Foreign tables to specify a SQL query to get remote data.
In our view, these facts limit the level of usability
these solutions can achieve that. Hence, we propose an architecture (and present its implementations) based on a hybrid conceptual model that enables users to write federated queries that are much richer and more comprehensive than the ones enabled by previous solutions.

Our solution encapsulates the remote data and the complexity of the underlying model. 
When accessing the data consumed and generated by the workflow execution, the user does not have to specify the paths to navigate to that data or to the remote data, since \hkpoly{} handles this task.
The user formulates a query considering a single repository (the \hkpolycatalog{}), and \hkpoly{} uses the (provenance) data references also stored in its catalog to get the remote data.

Giacomo et al.~\cite{degiacomo-etal:2016:ontologies-semantic-data-integration} present an overview of the concept ``Ontology-Based Data Access'' (OBDA), propose a general OBDA framework, and exemplify its instantiation theoretically. 
An OBDA system implements three components:
    (i) Ontology: provides a formal high-level representation of a domain, \eg{} used by clients to formulate queries;
    (ii) Data source layer: the remote data stores;
    (iii) Mappings: explicitly representations that map data sources and ontologies used to translate the client operations (\eg{} query answering) in terms of actions on the data sources.
The general approach of the framework is reasonably related to our proposal. 
However, it focuses on relational remote data stores.
The authors point out that the encapsulation of NoSQL databases is still a challenge. 
Besides, they do not present the framework implementation and experimental evaluation nor handle data linkage of data stores data. 
Our approach is broader, encompassing heterogeneous data models and tackling the data linkage problem.

Maccioni and Torlone~\cite{maccioni-and-torlone:2018:augmented-access-polystore} propose a data manipulation approach in polystores aka (query) augmentation which considers that objects in different data stores have a probabilistic relationship discovered by applying ML techniques. 
It allows query answering enrichment over a local database, using data from other databases inside a polystore system.
Although we share the data linkage concept to bind related data objects in a polystore system, our approach is quite different. 
While \cite{maccioni-and-torlone:2018:augmented-access-polystore} relies on learning through user's data exploration and crawling to understand how data objects are related, we provide data linkage using provenance besides a high-level view (the GCS) for users through integrating multiple databases.

%% file: body/sections/z-conclusion.tex
\section{Conclusion}
\label{sec:conclusion}

Provide an integrated view of heterogeneous data residing in different types of data stores is a big challenge~\cite{stonebraker:2015:polystore, ozsu-valduriez:2020:principles-of-data-systems}.
The main problem considered in this work is how to build data connections between such data and execute queries in an efficiently way.

The main contribution of this work is \hkpoly{}, a novel polystore architecture that uses domain ontology, remote data stores' schema metadata, data mappings, and knowledge graphs to support users with a single abstract global conceptual schema to write their queries, encapsulating data heterogeneity, location, and linkage.
As secondary contributions:
    (i) We analyzed the related components of federated database systems, including Multidatabase systems, SQL/MED specification, PostgreSQL FDW, Hyperknowledge, Polystore etc.;
    (ii) We presented an implementation of the architecture and we analyzed it in a real case within the O\&G industry;
    (iii) We demonstrated how a knowledge graph-centric approach could improve polystore queries by augmenting their semantics and facilitating the construction of queries that processes data in heterogeneous remote data stores linked by using provenance.

In our solution, a user application interacts with \hkpoly{} through its provisioned services.
\hkpoly{} supports the requirements presented in Section~\ref{sec:requirements}, such as creating domain models, loading remote data stores' schema metadata, and processing user queries. 
We presented the architecture and its implementation to meet such requirements with examples to visualize and illustrate the proposal.

We evaluated the query processing feature in which the input query is at the domain abstraction level.
The user does not have to know the complexities underlying the heterogeneous remote data systems.
Using \hkpoly{}, the user writes less complex queries. 
The evaluation in the O\&G scenario shows that the proposed architecture allows query writing that is two times less complex than the query one should write to use the multidatabase system (\eg{} \postgresqlfdw{}) directly.
Considering the processing time, \hkpoly{} adds an excess of no more than 30\% for transforming the high-level query (Query~\ref{lst:hyql-scenario}) in the \postgresqlfdw{} query (Query~\ref{lst:polystore-sql}).
Although the current experimentation deals with only one use case, we got promising results on a real scenario that demonstrate applicability and utility of the proposal,
It points out the use of schemas, provenance, and mappings among these data is a path to encapsulating the complexities of accessing and linking data from heterogeneous data stores.

In future work, we aim at evolving \hkpoly{} implementation to process other query operators and workflows of other scenarios with distinct structures besides improving the query execution processing time.
We also aim at handling multimedia data (\eg{} video, audio, and text)~\cite{pouyanfar2018multimedia} using the constructs of \hk{} that represent this kind of data.
In another direction, we can enhance the architecture by using ML techniques to augment polystore integration metadata like~\cite{maccioni-and-torlone:2018:augmented-access-polystore}.

%% file: body/sections/zz-appendix.tex
\appendix

\section{Appendix with more details about the paper published at SBSI 2024}

The following sections present details about the paper that was not included in the paper submission to SBSI due to lack of space.

\subsection{Use of the scientific research method}

We followed the scientific research method in this work executing the following steps:

\begin{enumerate}
    \item We defined the problem we are working on: querying heterogeneous data generated by business processes that are not explicitly connected.
    \item We searched and read works to identify solutions in the literature that tackle that problem. We found such a solution was an open issue.
    \item We defined the research question and used the Representation Theory to create our solution.
    \item We identified the requirements and designed the data structures and components of the architecture (i.e., the deep structure).
    \item We identified the requirements considering the literature on multi-database systems, and we identified how clients ("users") would interact with our architecture (i.e., the surface structure).
    \item We defined the metrics we would measure considering the user and database perspective to evaluate the proposal.
    \item We found a scenario where the problem we were tackling held.
    \item We ran experiments considering varying data volumes to analyze the database perspective, measure the processing time, and compare the processing time against PostgreSQL FDW.
    \item We analyzed the query complexity from the user's perspective by counting the query components to estimate the user's cognitive load. We compared the query written using the query language employed by HKPoly against the corresponding query in SQL.
\end{enumerate}

\subsection{Use of Representation Theory}

The Representation Theory was used to design the architecture, detail its components, and how they interact. Following the theory concepts, we presented the requirements and developed the data structures and components of the architecture (i.e., the deep structure). Afterward, we detailed how clients (``users'') interact with an architecture implementation (i.e., the surface structure). We also presented an exemplary architecture implementation, which we named HKPoly. We give details about its implementation, illustrating how it implements the proposed components and accesses the knowledge graph (i.e., physical structure).

Other IS theory could also strengthen the proposal construction, like Organizational Information Processing Theory and General Systems Theory.

\subsection{How the work was structured}

In the Introduction, we present the motivation, problem, gaps in the current solution, the research question, the goal of our work, and our proposal. We explain the background required to understand our work in Section 2, i.e., before presenting the solution details. In Section 3, we present the requirements a solution should support to tackle the research problem and our proposal of architecture to support them. The architecture is presented without implementation details so that different technologies can be used for one who wants to develop a solution following our architecture proposal. Afterward, in Section 4, we present an implementation of our architecture proposal considering the state-of-the-art technologies and a scenario to illustrate its use. We also present an evaluation of the implementation to demonstrate its feasibility, considering query complexity from the user and database perspective. After presenting the details of our solution and its implementation, we believe the reader has the knowledge about it and can understand the comparison we do to existing works – presented in Section 5. Finally, we present our conclusions, highlight the contributions, and present proposals for future work in Section 6.

\subsection{Comparison against big data processing systems concerning the provenance aspect}

To the best of our knowledge, most big data processing systems do not support provenance data management natively, relying on "plugins" to add such functionality. For example, work [1] adds provenance to the Apache Pig tool [2], and work [3] adds provenance to Spark [4]. However, none of these provenance plugins, alongside big data tools, deal with provenance in the form of knowledge graphs. Furthermore, as several of these plugins do not follow the W3C PROV standard, interoperability with other provenance databases is complex, which is also one of the essential aspects of our approach. As for data audit tools, they are typically considered complementary, not competitors, to provenance data management tools.

[1] Amsterdamer, Yael et al. “Putting Lipstick on Pig: Enabling Database-style Workflow Provenance.” Proc. VLDB Endow. 5 (2011): 346-357.

[2] https://pig.apache.org/

[3] Interlandi, Matteo et al. "Adding data provenance support to Apache Spark." The VLDB journal: very large databases: a publication of the VLDB Endowment vol. 27,5 (2018): 595-615. doi:10.1007/s00778-017-0474-5

[4] https://spark.apache.org/

\subsection{Other papers of SBSI related to this work}

Due to lack of space we could not include in the paper other references that are related to our work, such as:

\begin{itemize}
    \item \cite{VillacaEtAl:2023:EPACompArchitecturalModel} L. H. N. Villaça, S. W. M. Siqueira, and L. G. Azevedo, "EPAComp: An Architectural Model for EPA Composition," in Proceedings of the XIX Brazilian Symposium on Information Systems, in SBSI '23. New York, NY, USA: Association for Computing Machinery, Jun. 2023, pp. 61–69. doi: 10.1145/3592813.3592889.
    
    \item \cite{CamposEtAl:2023:OntologybasedDataIntegration} J. G. Campos, V. P. De Almeida, E. M. De Armas, G. M. H. Da Silva, E. T. Corseuil, and F. R. Gonzalez, "INSIDE: an Ontology-based Data Integration System Applied to the Oil and Gas Sector," in Proceedings of the XIX Brazilian Symposium on Information Systems, in SBSI '23. New York, NY, USA: Association for Computing Machinery, Jun. 2023, pp. 94–101. doi: 10.1145/3592813.3592893.
    
    \item \cite{VillacaEtAl:2020:MicroserviceArchitectureMultistorea} L. H. N. Villaça, L. G. Azevedo, and S. W. M. Siqueira, "Microservice Architecture for Multistore Database Using Canonical Data Model," in Proceedings of the XVI Brazilian Symposium on Information Systems, in SBSI '20. New York, NY, USA: Association for Computing Machinery, Nov. 2020, pp. 1–8. doi: 10.1145/3411564.3411629.
    
    \item \cite{MendesEtAl:2019:PolyflowSOAAnalyzing} Y. Mendes, R. Braga, V. Ströele, and D. de Oliveira, "Polyflow: A SOA for Analyzing Workflow Heterogeneous Provenance Data in Distributed Environments," in Proceedings of the XV Brazilian Symposium on Information Systems, in SBSI '19. New York, NY, USA: Association for Computing Machinery, May 2019, pp. 1–8. doi: 10.1145/3330204.3330259.

\end{itemize}